\documentclass[sn-mathphys]{sn-jnl}

\usepackage{natbib}

\jyear{2023}%

\begin{document}

\title[Study of scalar resonance in the $\omega\phi$ system]{
Study of a near-threshold scalar resonance in the $\omega\phi$
system in pion-Be interaction at momentum of 29 GeV
}



\author[1]{\fnm{V.A.} \sur{Dorofeev}} 
\author[1]{\fnm{D.R.} \sur{Eremeev}}
\author[1]{\fnm{V.G.} \sur{Gotman}} 
\author[1]{\fnm{A.V.} \sur{Ivashin}}

\author[1]{\fnm{I.A.} \sur{Kachaev}}
\author*[1,2]{\fnm{Yu.A.} \sur{Khokhlov}}\email{Yury.Khokhlov@ihep.ru}
\author[1]{\fnm{M.S.} \sur{Kholodenko}}
\author[1]{\fnm{V.F.} \sur{Konstantinov}}
\equalcont{Deceased}

\author[1]{\fnm{V.I.} \sur{Lisin}}
\author[1]{\fnm{V.D.} \sur{Matveev}}
\author[1]{\fnm{E.V.} \sur{Nazarov}}
\author[1]{\fnm{V.I.} \sur{Nikolaenko}}
\author[1]{\fnm{A.N.} \sur{Plekhanov}}
\author[1]{\fnm{D.I.} \sur{Ryabchikov}}
\author[1]{\fnm{A.A.} \sur{Shumakov}}
\author[1]{\fnm{V.P.} \sur{Sugonyaev}}
\author[1,2]{\fnm{A.M.} \sur{Zaitsev}}

\affil[1]{ 
\orgname{NRC "Kurchatov Institute"- IHEP}, \orgaddress{\street{Nauki Sq. 1}, \city{Protvino}, \postcode{142281}, \state{Moscow region}, \country{Russia}}}

\affil[2]{
\orgname{MIPT}, \orgaddress{\street{Institutsky Lane 9 }, \city{Dolgoprudny}, \postcode{141701}, \state{Moscow region}, \country{Russia}}}



\abstract{
The charge-exchange reaction 
$\pi^- \text{Be} \rightarrow \text{A}\,\omega(782)\phi(1020)$ with  
$\omega \rightarrow \pi^+\pi^-\pi^0$ and $\phi \rightarrow K^+K^-$  
is studied with the 
upgraded VES facility (U-70, Protvino) using  a 29 GeV pion  beam.
The distribution of the $\omega\phi$ invariant mass  shows a near-threshold  enhancement. 
Partial-wave analysis reveals that an  isoscalar scalar state with $J^{PC}=0^{++}$  dominates in this mass region. 

Considering  the observed signal as an $f_0$ resonance, and using the one-pion-exchange
model, the product  of the branching fractions into two channels $Br(f_0\rightarrow \pi\pi) Br(f_0\rightarrow \omega\phi)$ is measured. Based on this value,  a large partial width is found for the radiative decay of $J/\psi$ into this state, which 
suggests a significant  glueball component.

The result is discussed  under  alternative assumptions concerning an identification of the observed signal as   $f_0(1710)$ or  $f_0(1770)$.}




\maketitle

\section{Introduction}\label{sec1_intr}

The sector of scalar mesons holds a unique position  in light-meson spectroscopy due to  the uncertainty 
and redundancy in assigning the observed states to the quark-model SU(3)$_\text{flavor}$ nonets.

Resonance structures  with quantum numbers  $I^GJ^{PC}=0^+0^{++}$  are observed in several reactions  in the mass range from  1700 to 1800 MeV. The Particle Data Group (PDG) \cite{PDG} combines  most of these observations   into a single and established resonance, known as the  $f_0(1710)$,    with  parameters   
 \begin{equation}
M=1733^{+8}_{-7} \, \text{MeV} , \quad \Gamma = 150^{+12}_{-10} \, \text{MeV} \quad .
  \label{f0_params}
 \end{equation}
  Ref.~\cite{PDG} also contains a review  of the experimental data and their  interpretations.
 
 The $f_0(1710)$  has been extensively studied in 
 $J/\psi$ radiative decays, with  peaks  observed in  the  $4\pi$ \cite{4pi}, $K\bar K$ \cite{KK1, KK2}, $\pi\pi$ \cite{pipi}, and $\omega\phi$ \cite{BES:2006vdb, BESIII:2012rtd} channels.   The $\omega\omega$ channel also shows   a $0^+0^{++}$  state  in this mass region \cite{omom}, which  can  be associated with the $f_0(1710)$ as well. 
The partial width 
$$
Br(J/\psi\rightarrow \gamma f_0(1710) 
_{\hookrightarrow (4\pi,\,K\bar K,\,\pi\pi,\,\omega\phi,\,\omega\omega)})
$$
is significantly  larger than the partial  widths for the $J/\psi$ radiative decays to other scalars. This observation suggests that the $f_0(1710)$  may have a glueball nature     \cite{glueball}.  
  
Convincing evidence for two $f_0$ resonances with significantly  different masses and  branching fractions  to major  channels in the  1.7 to 1.8 GeV mass region  are presented in  Refs.~\cite{Bugg1, Ablikim1}.  The parameters of these   resonances were  determined  through a coupled-channel  analysis of  BESIII data in Ref.~\cite{Sarantsev:2021ein}.   

Our particular interest concerns the $\omega\phi$ channel. The BESII experiment  reported  in  Ref.~\cite{BES:2006vdb} 
a near-threshold peak with $J^{PC}=0^{++}$  in the $\omega\phi$  invariant mass spectrum in the reaction $e^+e^- \rightarrow
J/\psi \rightarrow \gamma \omega\phi$, which was later confirmed
by BESIII in Ref.~\cite{BESIII:2012rtd}. The VES experiment  observed a similar signal in the interaction of a pion beam with a beryllium target \cite{VES:2010uah}. 

The results presented in this paper are based on new data obtained with the upgraded VES setup and supersede the results of Ref.~\cite{VES:2010uah}. 
Our analysis employs the same partial-wave analysis technique as the preceding analysis, but uses an improved experimental setup with upgraded detectors, more detailed  and  accurate detector simulation, and more precise data. Additionally, we  performed   a cross-section  measurement. This   enables us to  compare our result  to  data from $J/\psi$ radiative decays and  theoretical models.  

This paper has the following structure:
in Sec.~\ref{sec2_ves_setup}, we give a description of the detector.  
Sections \ref{sec3_ev_select} and \ref{sec_om_phi} describe the selection procedure   and   some general features of  the  reaction under study.    
In Sec.~\ref{sec4_pwa},  we present  the partial-wave analysis (PWA) of the $\omega\phi$ system. 
The results of the PWA  are  discussed  in Sec.~\ref{sec6_results}.  The  signal  observed  at the threshold in the  scalar wave   of  the $\omega\phi$ system is similar to the signal  in $J/\psi$  radiative decay.    
To compare  this  signal  in the two production reactions,  we  estimated   the ratio of the  scalar-wave intensities  from   our  analysis  and  
 a previous VES analysis of the $\omega\omega$    channel  \cite{VES:2010uah,bib_om_om}. The    ratio we  find   is consistent    with the ratio observed in   $J/\psi$  decays. Therefore,   we   conclude that   the   object  observed  in the charge-exchange reaction is the same as the one  observed  in $J/\psi$ radiative  decays, which is  assigned  to the $f_0(1710)$  by the PDG.   
 Applying  the one-pion-exchange approximation to our data,  we  estimate  the product of the  branching fractions  into $\pi\pi$ and $\omega\phi$ for  the studied  $f_0$ state. 
In the discussion of our result,  we use two alternative assumptions: we identify the $f_0$  either with the  $f_0(1710)$  with parameters given in Eq.~(\ref{f0_params}), or  with the  $f_0(1770)$ from Ref.~\cite{Sarantsev:2021ein} with parameters 
 \begin{equation}
M= (1765 \pm 15) \, \text{MeV} , \quad \Gamma = (180 \pm 20) \, \text{MeV} \quad .
  \label{f01770_params}
 \end{equation}
 For both  cases,  we calculate the branching fraction for
$J/\psi \rightarrow \gamma f_0$ and compare it  with theoretical estimates  for the radiative $J/\psi$  decay to a glueball. 
Our findings are summarised in Sec.~\ref{sec_conclusion}. 

\section{VES Setup}\label{sec2_ves_setup}

\begin{figure}[H]
  \centering
  \includegraphics[width=0.9\textwidth]
  {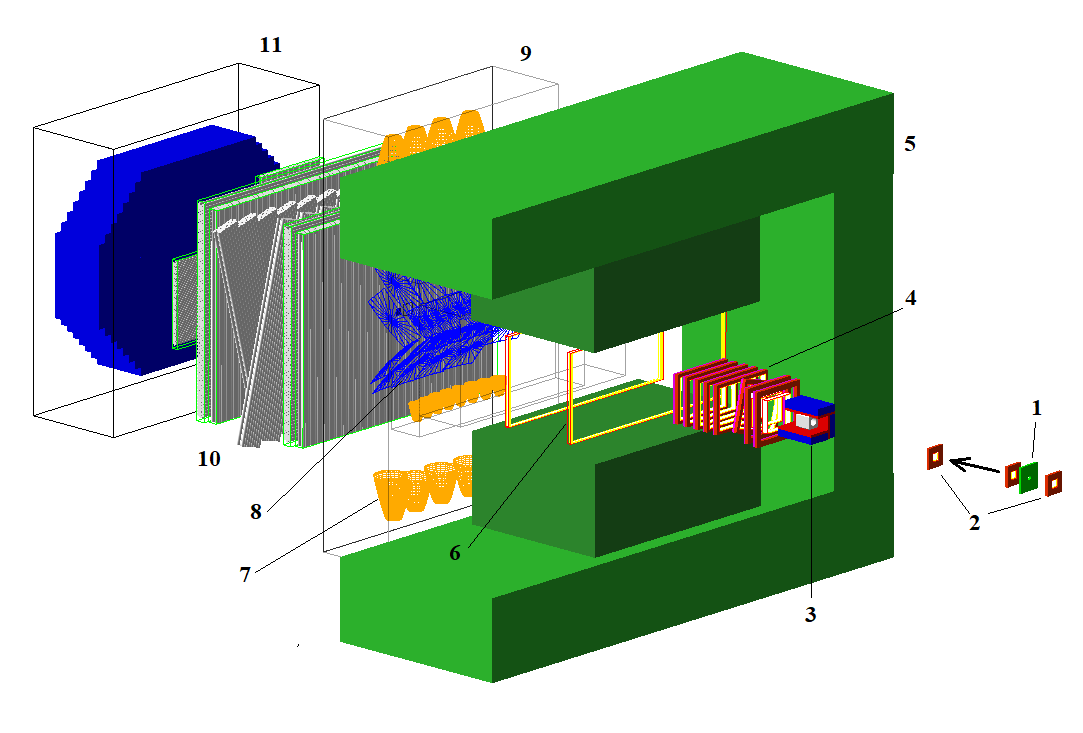}
  \caption{Schematic view of the VES setup with the  downstream  counter of the beam telescope (1); the beam proportional chambers  (2);  the veto box with the target (3); the proportional chambers of the spectrometer (4); the spectrometer magnet (5); the drift chambers (6);  the light-collecting cones (7), the mirrors (8), and the body (9) of the Cherenkov counter; the drift-tube  stations (10); and the electromagnetic calorimeter (11). The arrow indicates  the  direction of the beam.  }
  \label{fig_VES_setup}
\end{figure}

The VES  fixed-target experiment is 
a wide-aperture magnetic spectrometer with electromagnetic calorimetry and particle  
identification (see Fig.~\ref{fig_VES_setup}). The setup has been in operation since 1988 \cite{VES-1} and  is located at a 
secondary beam line of the U-70 proton synchrotron in Protvino.  
About a  decade ago, the facility  was significantly upgraded, while retaining  the general structure. The electromagnetic calorimeter and a major part of the tracking detectors were replaced. The  particle identification   was improved. The setup was equipped with a new data acquisition  and  trigger system. These upgrades help  to suppress backgrounds, increase the  rate of data collection, extend the list of measurable  reactions,  and allow us to measure their cross sections.   

The data used in  this  
analysis were obtained with a nominal beam momentum of 29 GeV and a typical momentum spread of $\pm 1\%$.  The  
intensity of the negatively charged particles delivered to the experiment  was typically (1 to 2)$\cdot 10^6$/s. The beam was composed  of approximately  98.0\% $\pi^-$, 1.6\% $K^-$, and 0.2\% 
$\bar p$. The remaining  0.2\% were  muons and electrons.
    
The beam section of the experiment (not shown in Fig.~\ref{fig_VES_setup}) consists of  Cherenkov threshold counters for the  identification of incoming pions, kaons, and antiprotons, beam-defining scintillators,  and proportional chambers,  which are used mainly for the  beam particle tracking. Together with the last dipole magnet of the beam line  they are also used to measure the beam momentum   $P_\text{beam}$.   The 1\%  accuracy of this measurement   gives us   limited control over   nucleon excitations  in the  target-fragmentation kinematic region using  the missing-mass technique.      
  
\begin{figure}[H]
\begin{minipage}{0.48\textwidth}
\hspace*{-1.1cm}\includegraphics[width=1.35\textwidth]{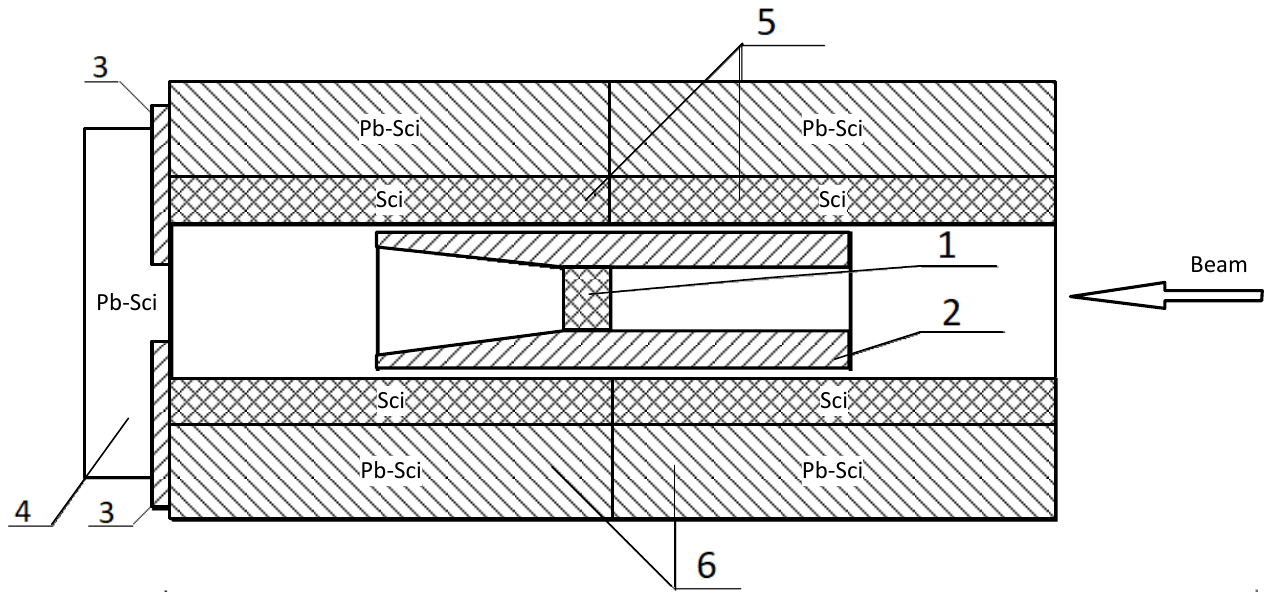}  
  \caption{Side  view of the target region (not to scale),  with the target (1); the holder (2); the lead convertor (3); the forward  sandwich detectors (4); the side scintillators  (5), and  the side sandwich detectors (6).}
  \label{target}
\end{minipage}
\hfill
\begin{minipage}{0.48\textwidth}
  \centering
  \includegraphics[width=1.0\textwidth]{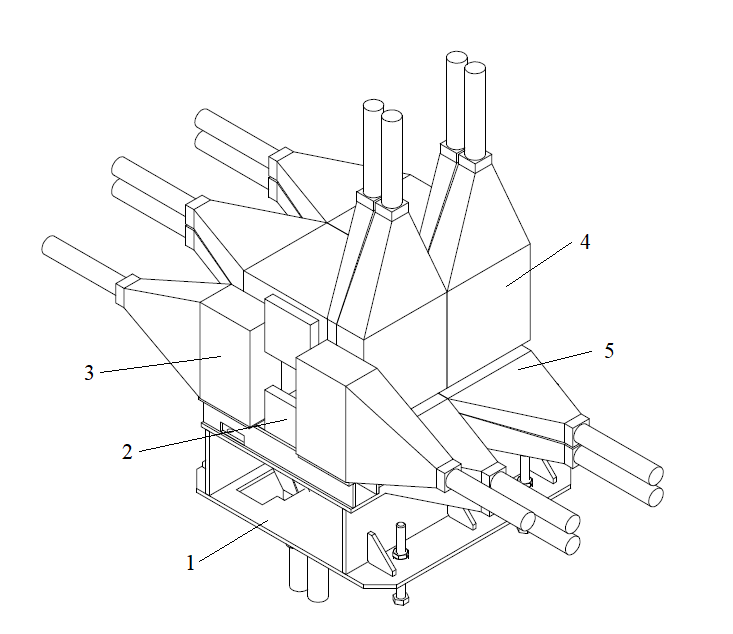}
  \caption{The veto box surrounding  the target, which includes the support (1),  lead converters (2),  forward sandwich detectors (3),  side sandwich detectors (4), and side scintillators (5). }
  \label{veto}
\end{minipage}
\ 
\end{figure}

The  beryllium  target has a diameter of 45 mm, which matches  the beam spot size, and a thickness of 40 mm, which corresponds to about 10\% of  radiation and  nuclear interaction lengths. The target  is inserted into  a hole  in an  aluminum holder (Fig.~\ref{target}). The holder  is a square prism of  dimensions   $12\times 12\times 20$ cm$^3$. It absorbs  slow recoiling protons and delta electrons.  The  hole in the holder has a cylindrical shape on the upstream side and a conical shape
on the downstream side.  

The target holder is  surrounded on four sides by two layers of  veto detectors, as shown in  Fig.~\ref{veto}.   The  inner layer is composed of plastic scintillators, while    the outer layer is made of
Pb-scintillator sandwich detectors.   The veto box is completed by two forward sandwich counters and a lead converter.  
The veto box  rejects events  where the produced particles leave the target at large angles.  The trigger uses the scintillation-counter signals, while the signals from the sandwich detectors  are digitized for  offline analysis.  

The veto  box with the target  is installed upstream of the main spectrometer magnet.   
This  dipole magnet has an aperture that is   about 2 m wide and   1 m high. The conical hole  of the target holder and the veto counters  match   this aperture. At a current of 4 kA,  the  integral of the vertical field component is approximately  1.5 Tm.  
   
Two 70 mm diameter scintillation disks, known as "Beam Killers" BK1 and
BK2, are positioned downstream of the magnet along the nominal beam trajectory
 to suppress events without  beam interaction.  
  
The spectrometer's tracking system comprises  proportional and drift chambers. Specifically, there are  three 
two-plane multi-wire proportional chambers (MWPC) with an active area of approximately  $40\times 25 \,\text{cm}^2$ 
and five two-plane MWPCs with a $60 \times 40 \, \text{cm}^2$ active area at the entrance of the magnet. There are also two two-plane small-gap 
 drift chambers inside the magnet and six planes of drift-tube detectors downstream of the magnet. Each drift-tube detector has  dimensions of about $1.5 \times 2.0 \,\text{m}^2$ and   consists    of three layers  of  30 mm diameter drift tubes.  The 
momentum resolution of the spectrometer is about 0.5\% at 3 GeV and 1.2\% at 25 GeV.
   
A 28-channel Cherenkov counter (Ch-28) is used to identify charged final-state particles. 
The threshold momentum for  pions is about 3.5 GeV  (see Ref.~\cite{Kholodenko:2020dxg} for details).

The amount of material, that  final-state photons have to traverse
before reaching the electromagnetic 
calorimeter (EMC) \cite{Dorofeev:2016vsp} at the downstream end of the setup is less than 15\% of  
radiation length.  The parametrization of the  energy resolution of the EMC is expressed as a quadratic sum: $\sigma_E/E = 0.069/\sqrt{E/1~\text{GeV}} \oplus 0.027$ and is used in the 1C fit to the $\pi^0$ mass (see Sec.~\ref{sec3_ev_select}). The  spatial  resolution of  EMC in the 
central part is 3.2 mm at 5 GeV.

Other systems of the upgraded setup are described in Refs.~\cite{DAQ,DCS}.

\section{Event selection}
\label{sec3_ev_select}
The analysis is based on data collected  during four run periods between  2013 and 2015. About $2.3\cdot 10^{11}$   beam particles  
passed through the beam telescope during  the live time of the experiment.
Events for the reaction 
 \begin{equation}
  \pi^- \text{Be} \rightarrow \text{A}\ \omega(782)\phi(1020)   
  \label{reaction}
 \end{equation}
 with $\omega \rightarrow \pi^+\pi^-\pi^0$ and $\phi \rightarrow K^+K^-$ are selected.
Here, symbol  A represents  a final-state nucleus plus a recoil neutron that either remains  in the nucleus or leaves  it.

The event candidates for the exclusive event meson system are required to have four particle
tracks,   two with positive and two with negative  charge, and two or three photon  clusters in the EMC with  energies 
$E_\gamma>200$ MeV.   If the third cluster, which  is likely caused by noise in the EMC,  is present, it should not be used in  a $\pi^0$  candidate (see below) and  its energy must not exceed 500 MeV. 
 Noise clusters are $\gamma$-like signals in the EMC that are not related to a  beam-induced reaction.   They may be caused by $\delta$ electrons,  pile-up hits, or electronic noise. 

\begin{figure}[H]
\begin{minipage}{0.59\textwidth}  \centering
  \includegraphics[width=0.9\textwidth]
  {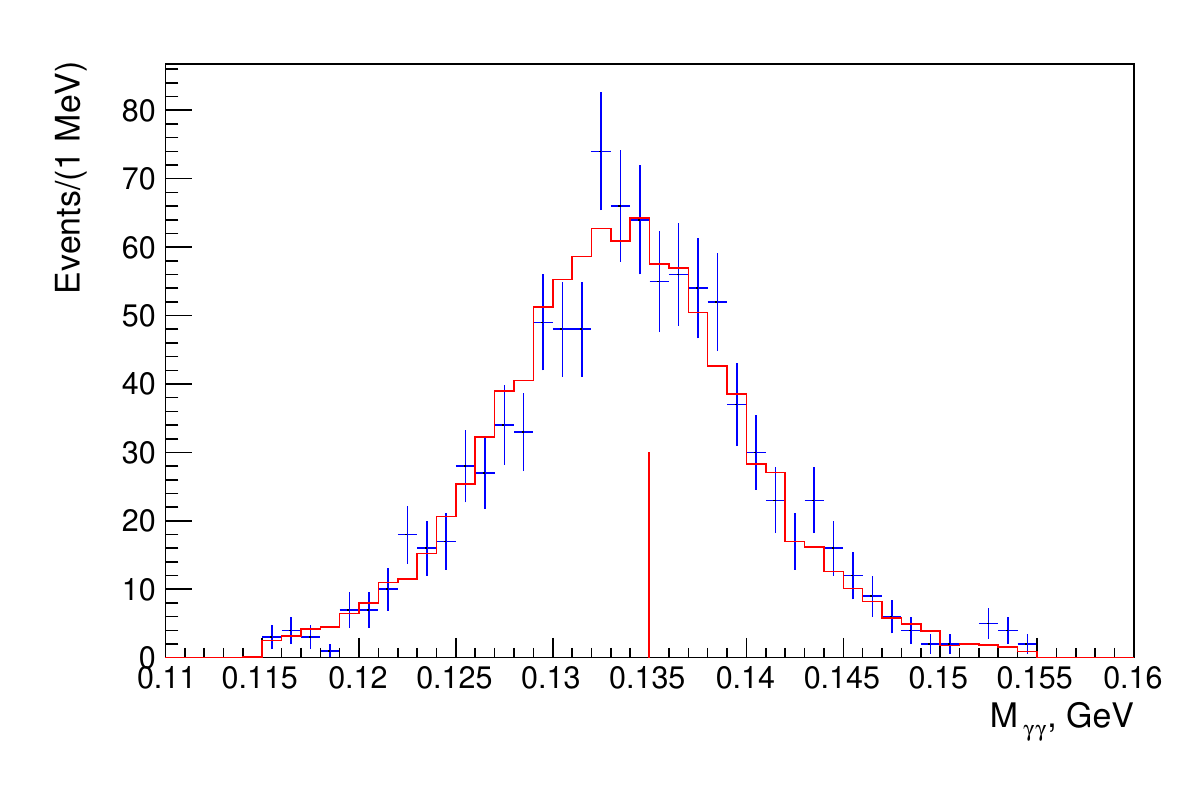}
  \caption{The $M_{\gamma\gamma}$ invariant mass spectra for   the selected measured  events (in blue)   and for the simulated data generated using the PWA result (in red).  The nominal $\pi^0$ mass is indicated by the vertical line.}
  \label{m2g}
\end{minipage}
\hfill
\begin{minipage}{0.4\textwidth}
  \centering
  \vskip -0.5cm
  \includegraphics[width=1.0\textwidth]{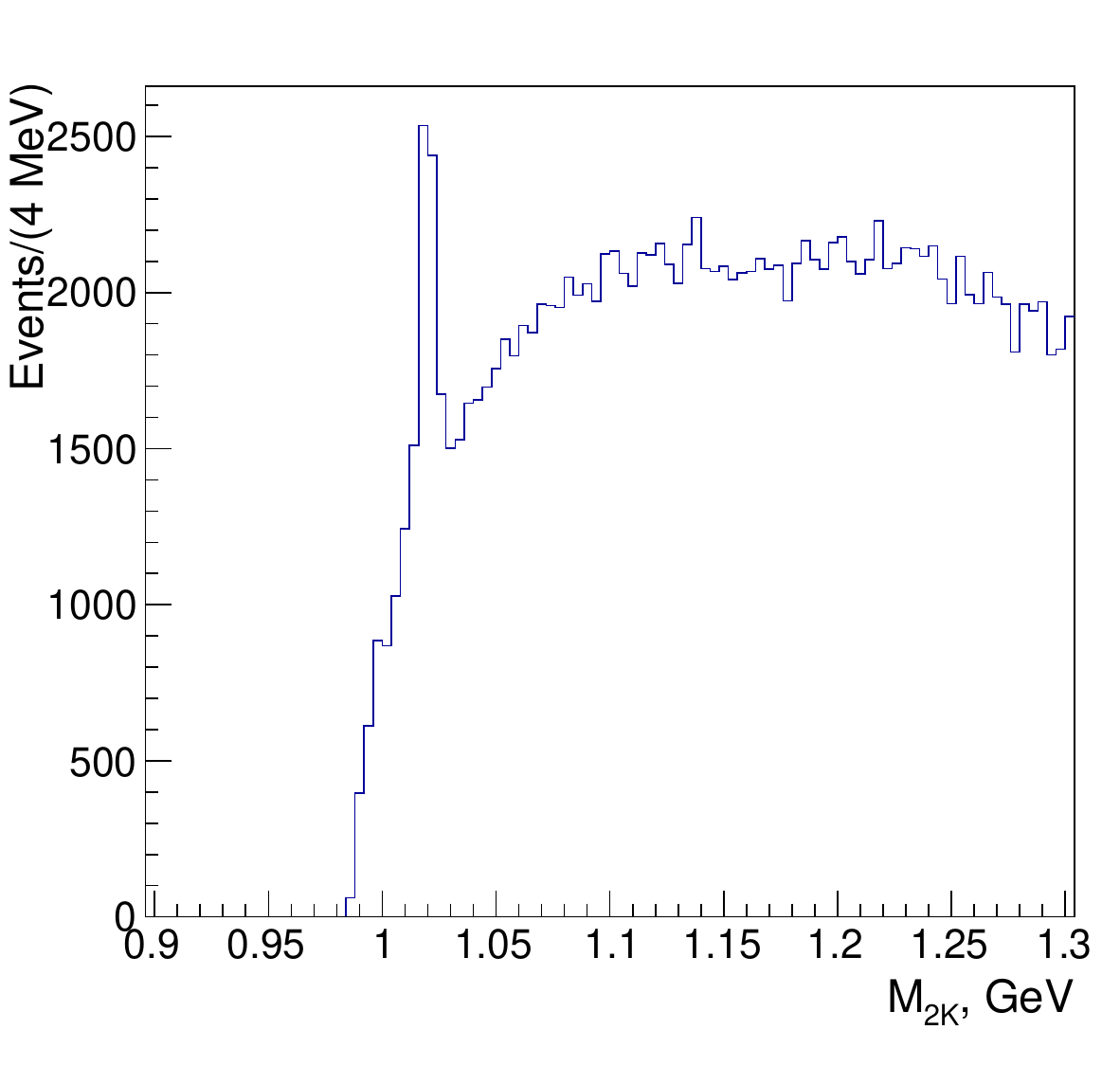}
  \caption{$K^+K^-$ invariant mass spectrum.}
  \label{fig_M_KK}
\end{minipage}
\hfill
\ 
\end{figure}

The distribution of the two-photon invariant mass $M_{\gamma\gamma}$ (Fig.~\ref{m2g}) 
exhibits a Gaussian peak from $\pi^0 \to \gamma \gamma$, which has a
width (RMS)  of 6.0 MeV and a peak position that deviates by about 1 MeV from the
nominal $\pi^0$ mass $M_{\pi^0}$.
 We select photon pairs  with  $\mid M_{\gamma\gamma} - M_{\pi^0} \mid< 20$ MeV as  $\pi^0$ candidates .  
  
To select  events where the beam $\pi^-$ interacted  with the Be target, the position of the reconstructed interaction vertex is required to be within the target volume.  The  vertex reconstruction provides a spatial  resolution of about  10 mm along the beam direction  
and of about 1 mm in the perpendicular plane. 

In the  momentum balance of the reaction, the recoil or potential target fragments can be omitted as the absolute value of the squared four-momentum $t$ transferred  from the beam to the target  is small (see below). Hence, the closeness of the total  momentum $P_\text{tot}$  of the detected forward-going  particles  to  $P_\text{beam}$  is a measure for the exclusivity.  The prominent peak in the $P_\text{tot}$  distribution (see Fig.~\ref{fig_filtered_ptot} and Sec.~\ref{sec_om_phi})  is selected to impose exclusivity, i.e. 
\begin{equation}
27.5 \,\text{GeV}< P_\text{tot}  < 31.0 \,\text{GeV} 
  \label{ptotcut}
  \end{equation}
  is required.
The tail at  low $P_\text{tot}$ values indicates  background form non-exclusive events.

Some events contain non-reconstructed large-angle tracks. These tracks appear as  
peripheral hits in the proportional chambers just after the target box. To enhance  the exclusivity of the sample, 
events with short tracks  are rejected.
  
Events where charged particles traverse either BK1 or BK2 are rejected at trigger level. 
In addition, none of the reconstructed tracks should pass through the circular regions, which are centred at the nominal positions of BK1 and BK2 and have a larger diameter of 80 mm. This takes into account potential counter inefficiencies, uncertainties in their positioning, and tracking resolution.

The Ch-28 detector is used in the threshold regime to   identify  charged particles.  Selected events have to have a higher likelihood for the $K^+K^-\pi^+\pi^-$  hypothesis than  for the  4$\pi^\pm$ hypothesis. Specifically, we limit the likelihood ratio to
$\mathcal{L}(2K2\pi)/\mathcal{L}(4\pi) > \alpha_0$. The parameter $\alpha_0$ controls both  the    efficiency and purity of the particle identification.  
We have chosen a value of $\alpha_0=2$ as a compromise between
enhancing the $\phi$ signal discrimination and maintaining event statistics.

\begin{figure}[H]
\begin{minipage}{0.42\textwidth}
  \centering
  \includegraphics[width=1.0\textwidth]{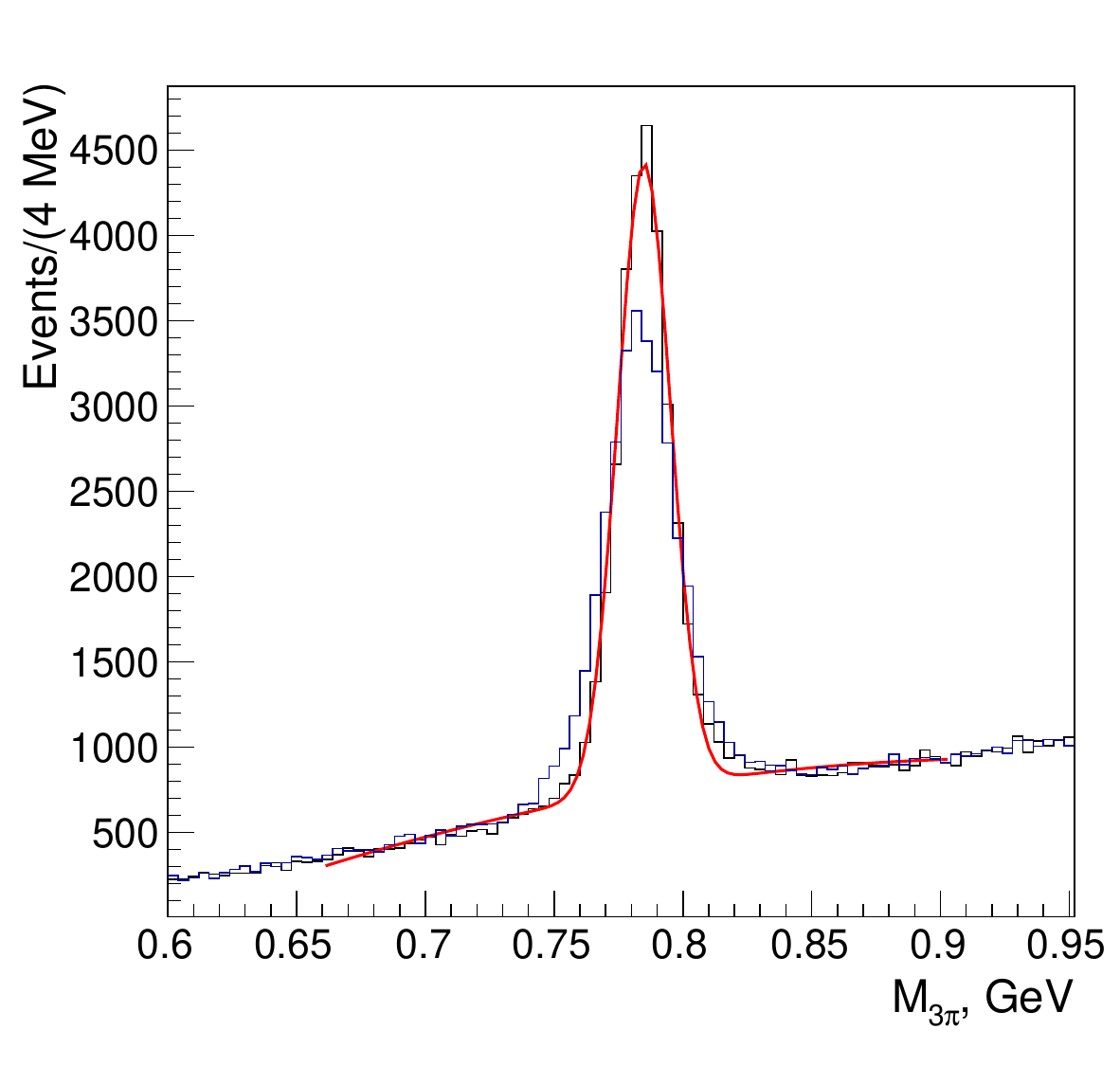}
  \caption{The invariant mass spectrum of $\pi^+\pi^-\pi^0$ is shown in the black and blue histograms, with and without the 1C fit to the nominal mass of $\pi^0$, respectively.
   The red curve  represents a fit with a sum of a Gaussian   and  a quadratic  function   for signal and background, respectively.}
  \label{fig_M_pipipi0}
\end{minipage}
\begin{minipage}{0.55\textwidth}
\centering
\vskip -1.cm
  \includegraphics[width=1.\textwidth]
  {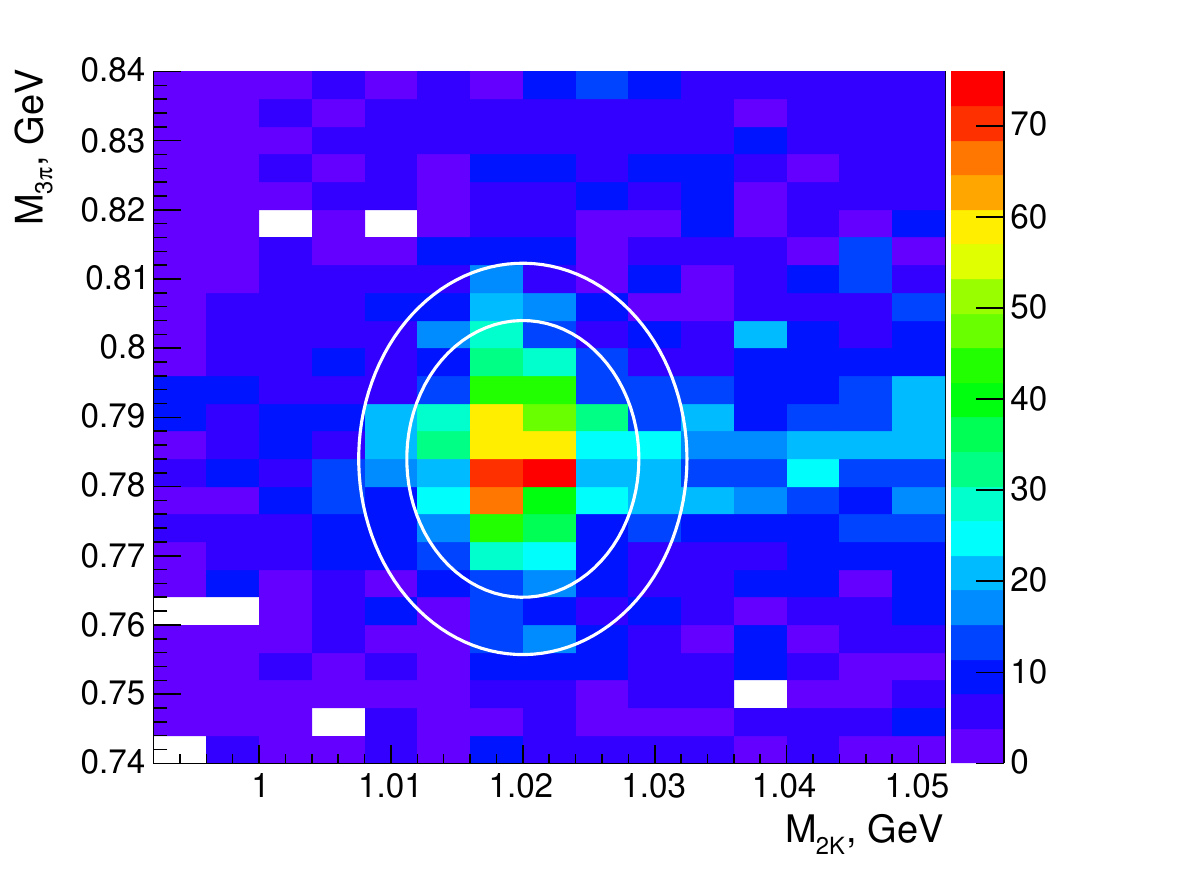}
  \caption{Two-dimensional distribution of the invariant masses for $\pi^+\pi^-\pi^0$  and  $K^+K^-$ . The  inner ellipse represents the signal region, while the elliptical ring corresponds to the background region.}
  \label{fig_M_KK_vs_M_pipipi0}
\end{minipage}
\end{figure}

The   invariant mass spectra for $K^+K^-$ and  $\pi^+\pi^-\pi^0$ in Figs.~\ref{fig_M_KK} and \ref{fig_M_pipipi0}  
show   clear  Gaussian-like peaks  for $\phi(1020)$ and $\omega(782)$ within 2 MeV of their nominal masses   \cite{PDG},
respectively. Both peaks sit on a  substantial background.  
We constrain the energies of the two $\gamma$ to the nominal $\pi^0$ mass in a 
1C kinematic fit, resulting in a  decrease of the Gaussian width  of the  $\omega$ peak from  about 13 to 10 MeV.

 Figure \ref{fig_M_KK_vs_M_pipipi0} shows the two-dimensional distribution of the $\pi^+\pi^-\pi^0$   and the $K^+K^-$ mass, with a clear signal for the associated production of $\omega$ and $\phi$  and less background than in  the one-dimensional projections shown in Figs.~\ref{fig_M_KK} and \ref{fig_M_pipipi0}.

The required signal events for reaction (\ref{reaction})  have to lie  within an  ellipse in the ($M_{2K}$, $M_{3\pi}$) plane.  This ellipse is defined by 
 \begin{equation}
r_{2M}^2 = \frac{(M_{3\pi} - M_\omega)^2}{\sigma_\omega^2} + \frac{(M_{2K} - M_\phi)^2}{\sigma_\phi^2} \leq 4 \, . 
\label{eq_ellipse}
\end{equation}
Here, $M_\omega$ and $M_\phi$ are the nominal masses of
$\omega(782)$ and $\phi(1020)$, respectively, taken from Ref.~\cite{PDG}, and 
 $\sigma_\phi = 4.4 \,\text{MeV}$ and   $\sigma_{\omega} = 10 \,\text{MeV}$ are  the RMS values for the corresponding Gaussian peaks.  The selected  region contains   1054 events.   

\section{General features  of the $\omega\phi$ system }\label{sec_om_phi}

\begin{figure}[H]
\begin{minipage}{0.47\textwidth}
\centering
\includegraphics[width=0.9\linewidth]{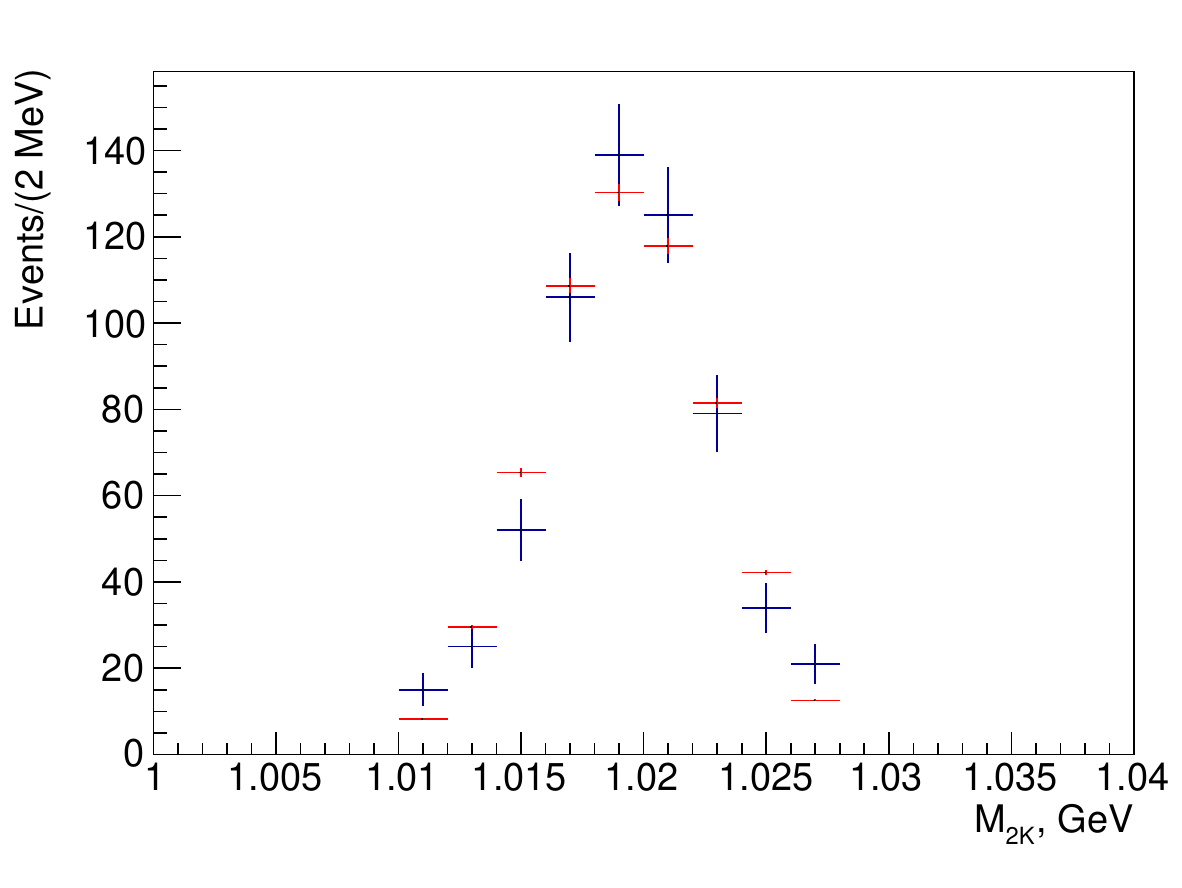}
\caption{$K^+K^-$  invariant mass spectrum  for   the selected measured  events (blue)   and for simulated data generated using the PWA result (red).}
\label{fig_predict_m_KK}
\end{minipage}
\hfill
\begin{minipage}{0.47\textwidth}
\centering
\includegraphics[width=0.9\linewidth]{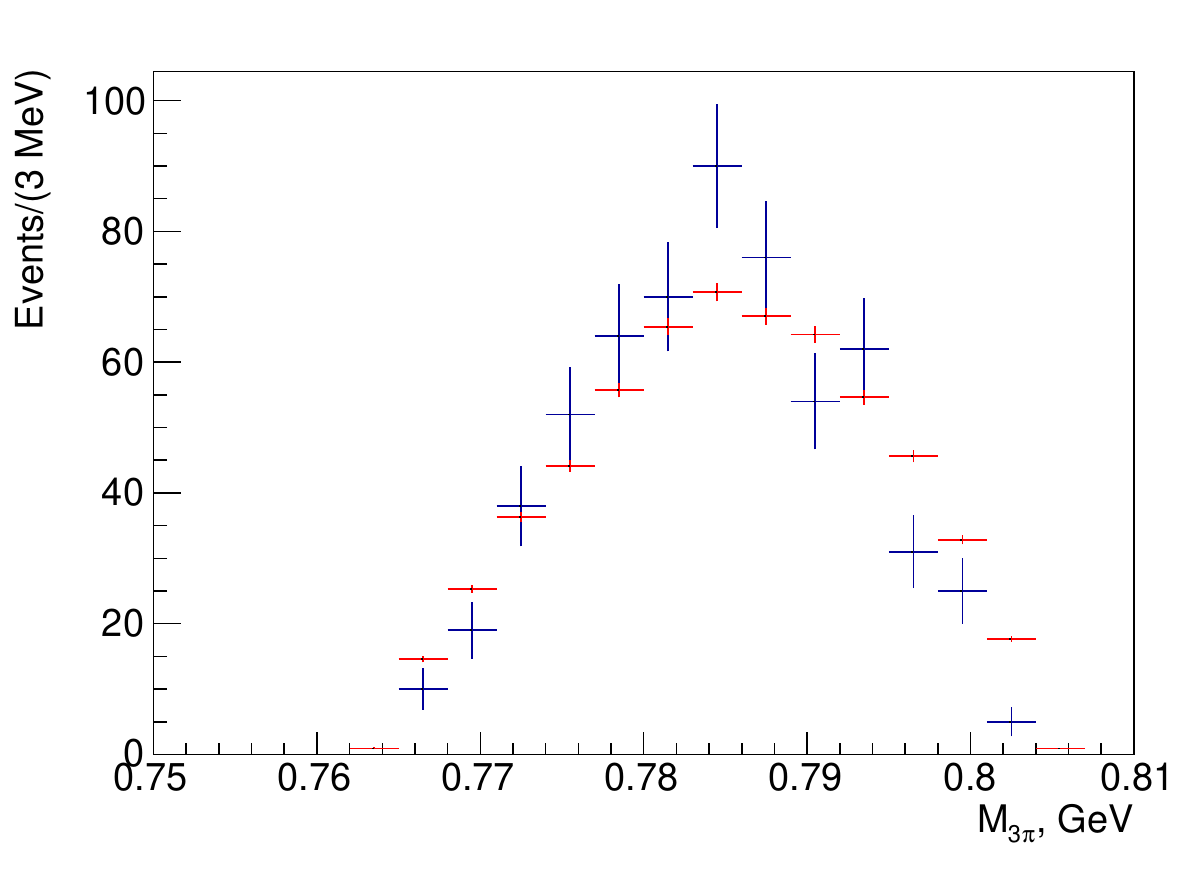}
\caption{$\pi^+\pi^-\pi^0$ invariant mass spectrum 
for   the selected measured  events (blue)   and for simulated data generated using the PWA result (red).}
\label{fig_predict_m_pipipi0}
\end{minipage}
\end{figure}

The $M_{2K}$ and $M_{3\pi}$  spectra for the selected events in the signal region defined by Eq.~(\ref{eq_ellipse}) are shown in Figs.~\ref{fig_predict_m_KK} and \ref{fig_predict_m_pipipi0}, respectively. For comparison, we superimpose the  distributions for simulated data generated using the PWA result
(see Sec.~\ref{sec4_pwa}).  

\begin{figure}[H]
  \begin{minipage}{0.50\textwidth}
  \centering
  \includegraphics[width=1.0\textwidth]{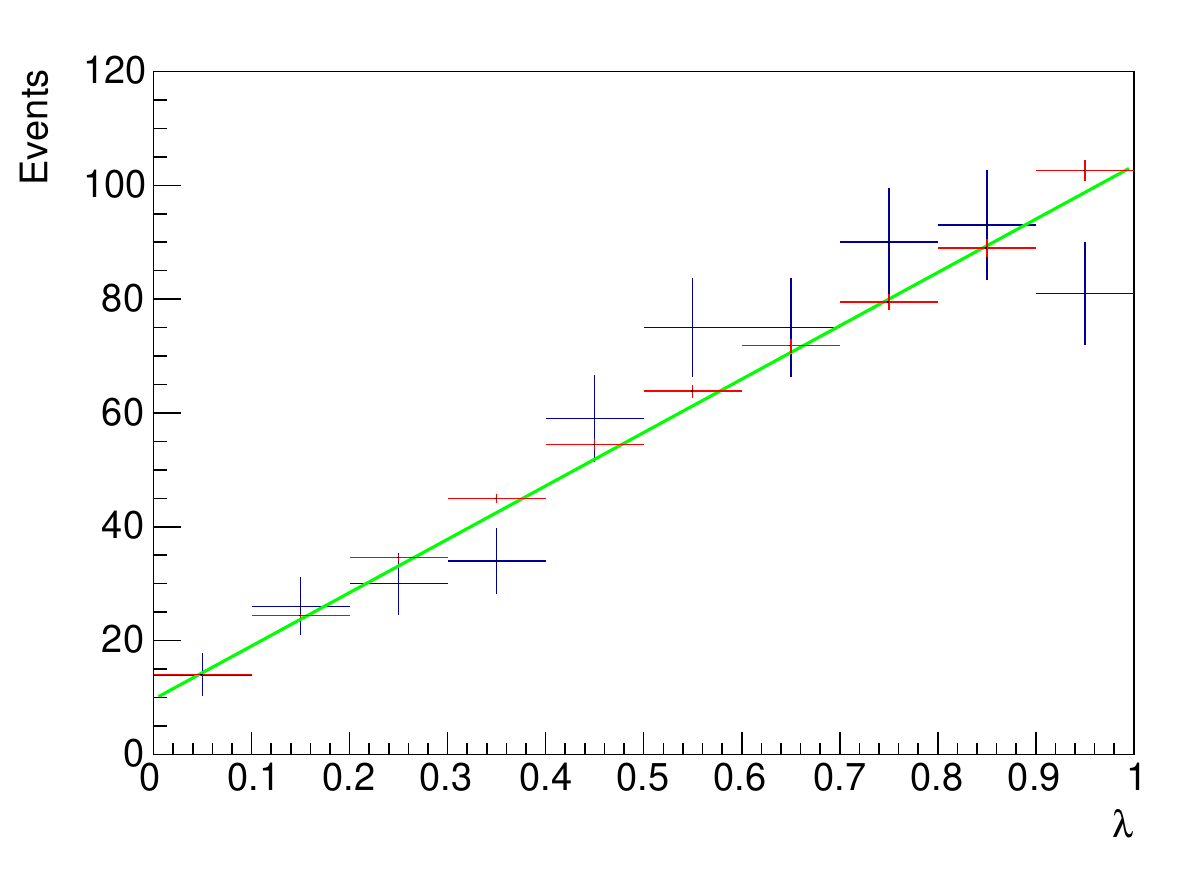}
  \caption{Distribution of the  kinematical variable $\lambda$ in Eq.~(\ref{eq_lambda}) for  the selected measured  events (blue)   and for simulated data generated using the PWA result (red).
   The  green line represents  a linear fit to the experimental data.}
  \label{fig_lambda}
\end{minipage}
\hfill
\begin{minipage}{0.45\textwidth}
  \centering
  \includegraphics[width=1.0\textwidth]
  {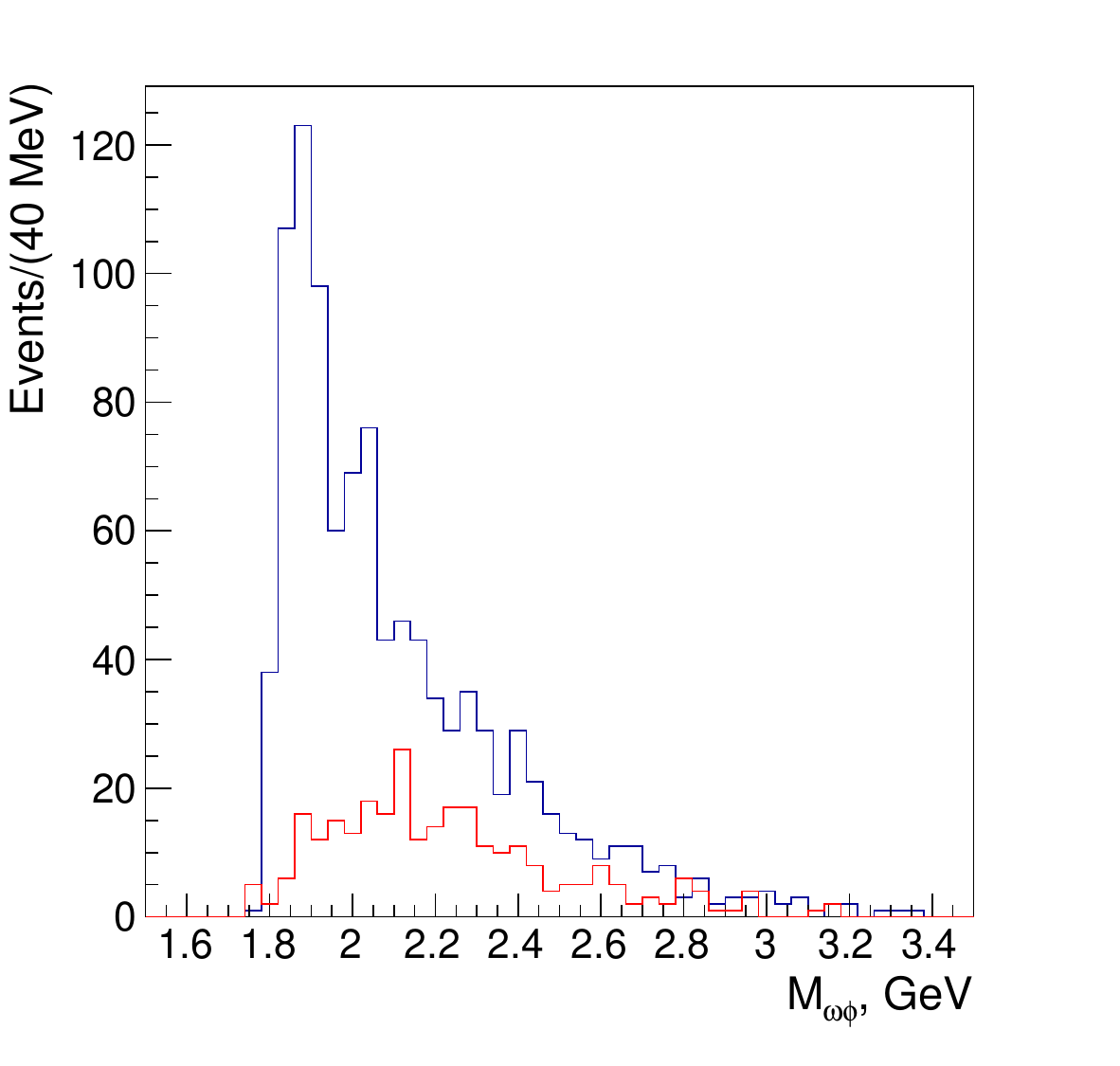}
  \caption{$K^+ K^-\pi^+\pi^-\pi^0$ invariant mass spectra in the signal region $r_{2M}^2 \leq 4$ (blue) and in the background region $4<r_{2M}^2 \leq 8$ (red).}
  \label{fig_M_KKpipipi0_bkg_ring}
\end{minipage}
\ 
\end{figure}

The $\omega\rightarrow 3\pi$ decay is characterized by a
linearly increasing   distribution of  events in the   kinematic variable  
\begin{equation}
\lambda = \frac{\mid\vec {p}_{\pi^-}\times\vec{p}_{\pi^+}\mid^2}{\lambda_\text{max}}\quad,
\label{eq_lambda}
\end{equation}
where  $\lambda_\text{max}=Q^2\large(Q^2/108 + m_\pi Q/9 + m_\pi^2/3\large)$ and $ Q=T_{\pi^+} + T_{\pi^-} + T_{\pi^0}$, with pion momenta $\vec{p}_\pi$ and kinetic energies $T_\pi$  in the $3\pi$ center-of-mass system (c.m.s.). The distribution of the experimental events is  shown in  Fig.~\ref{fig_lambda}. The constant component   is mainly due to the non-$\omega$  background. Its  contribution  is estimated from a  linear fit to be  about 20\%. 

The spectrum of the $K^+ K^-\pi^+\pi^-\pi^0 $ invariant mass   $M_{\omega\phi}$  is sharply peaked  near  the threshold (blue histogram in Fig.~\ref{fig_M_KKpipipi0_bkg_ring}).  This peculiarity is the main focus of our study. 
In contrast, the  $M_{\omega\phi}$ spectrum  in the elliptical region   $4 < r_{2M}^2 \leq 8$, which contains mostly background (see Fig.~\ref{fig_M_KK_vs_M_pipipi0}), does not show this structure (red histogram in Fig.~\ref{fig_M_KKpipipi0_bkg_ring}). 

\begin{figure}[H]
\begin{minipage}{0.45\textwidth}
\centering
\vskip -1cm 
\includegraphics[width=1.\textwidth]
{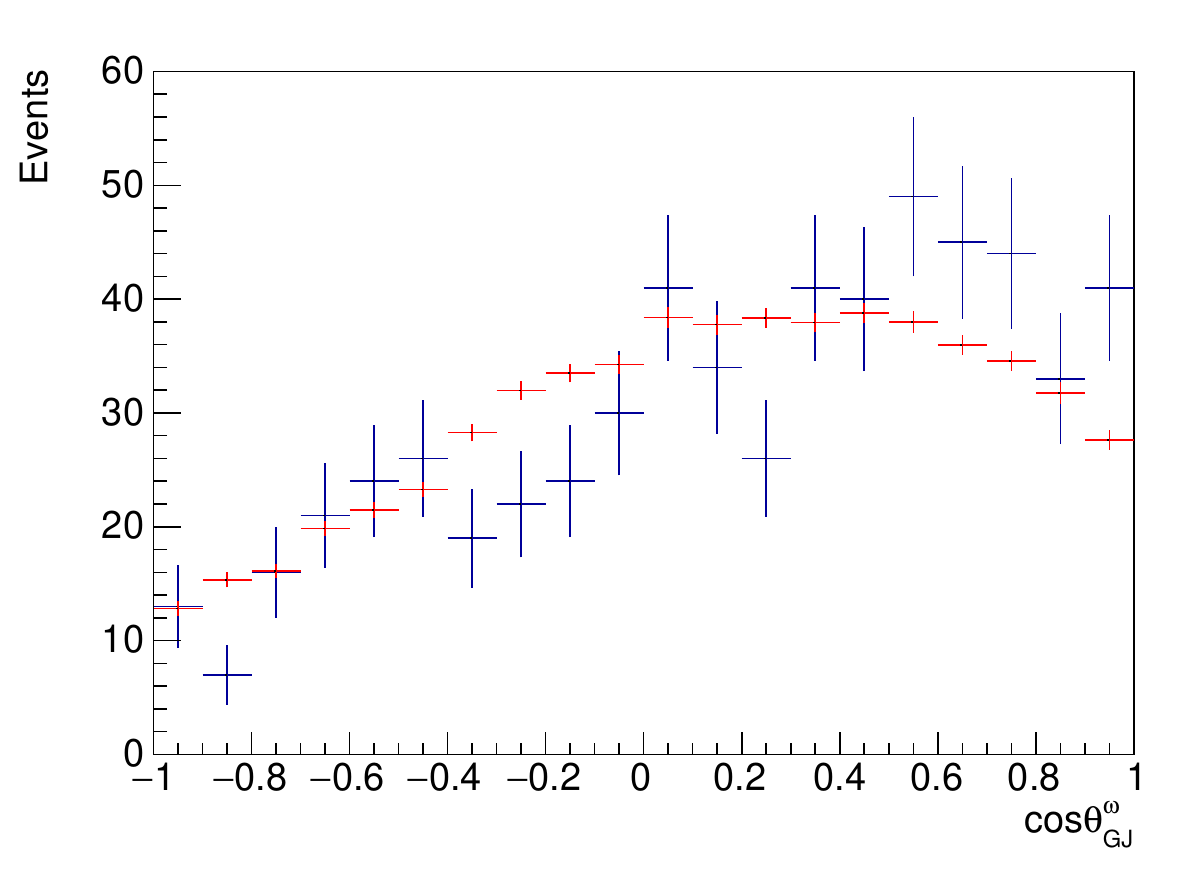}
\caption{Distribution of the cosine of the   Gottfried-Jackson angle of  the $\omega$ momentum in the $\omega\phi$ rest frame  for   the selected measured  events (in blue)   and for simulated data generated using the PWA result (in red).
}
\label{cosGJ_om}
\end{minipage}
\begin{minipage}{0.45\textwidth}
\centering
\includegraphics[width=1.\textwidth]{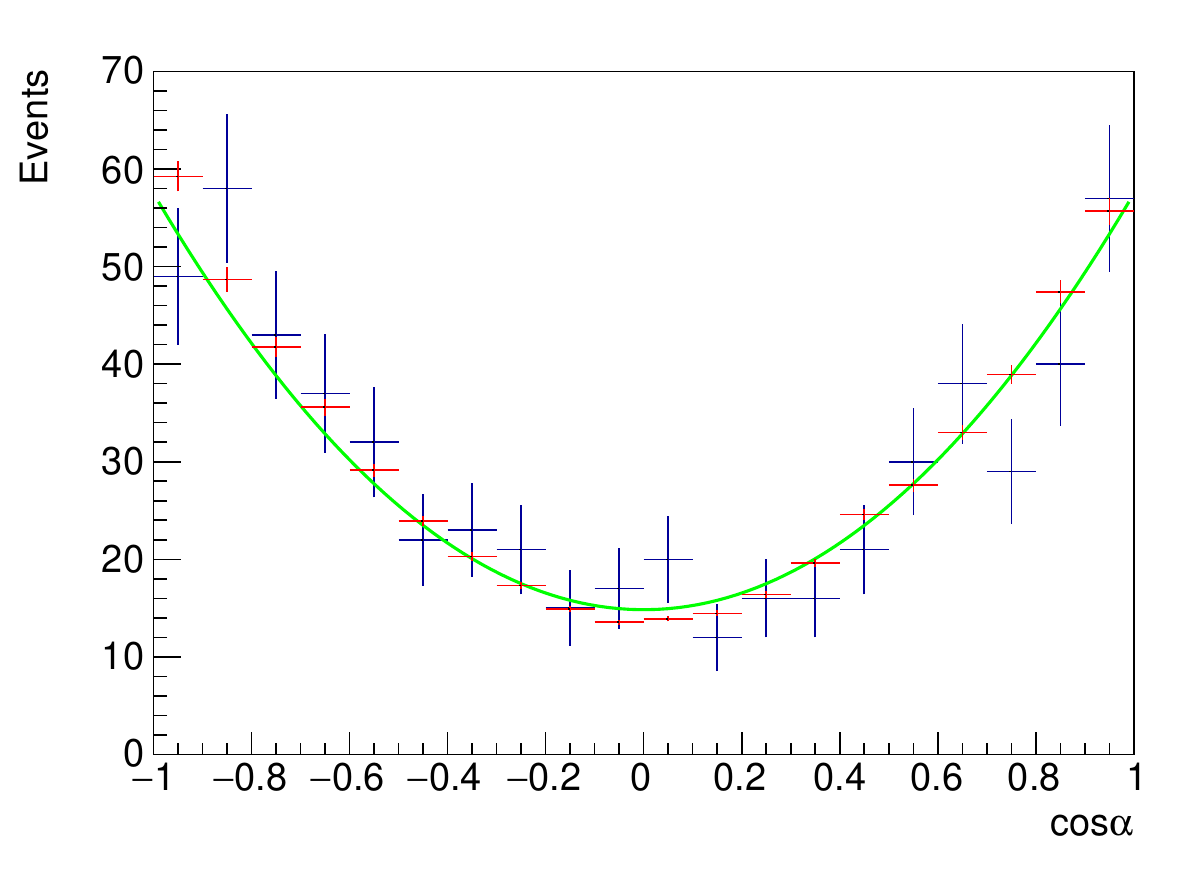}
\caption{Distribution of the cosine of the angle $\alpha$ between the normal  of the $\omega$ decay plane in its rest frame and the $K^+$ momentum in the $\phi$ rest frame for   the selected measured  events (in blue)   and for simulated data generated using the PWA result (in red).
 The green curve  represents a quadratic function  fitted to the experimental data. }
\label{fig_angle_om_kaon}
\end{minipage}
\end{figure}

The distribution of the cosine of the   Gottfried-Jackson angle of the $\omega$ momentum in the $\omega\phi$ rest frame is shown in Fig.~\ref{cosGJ_om}. 
Its shape is strongly affected by the efficiency.   
In contrast, Fig.~\ref{fig_angle_om_kaon} shows the distribution of the cosine of the angle $\alpha$ between the normal of the $\omega$ decay plane in its c.m.s. and the $K^+$ 
momentum in the $\phi$ c.m.s., which  is  less affected by the efficiency. This distribution  has a distinct  parabolic-like component $\propto\cos^2 \alpha$, as expected 
for the decay of a spinless particle to two vector particles with an  orbital angular momentum of $L=0$. Therefore, this distribution indicates a significant contribution from  a  $J^{PC}=0^{++}$ state.

The elliptical cut described in Eq.~(\ref{eq_ellipse})  results in some background being present below  the two-dimensional   $\omega\phi$ peak.  In the partial-wave analysis outlined  in Sec.~\ref{sec4_pwa},  this background is not subtracted but  is instead effectively  accounted for in the PWA model. 
In this section, we present   kinematic 
 distributions where we subtracted this background.  
The non-$\omega$ background is removed through bin-by-bin subtraction applied to the sample without the cut in Eq.~(\ref{eq_ellipse}). 
 For each bin of the variable in question, we require  $M_{2K}$ to lie within the $\phi$ band and fit 
the uncut $M_{3\pi}$ distribution  with the sum of a linear background and  a  Gaussian signal. In this fit,   the Gaussian parameters are fixed to the  $M_\omega$ and  $\sigma_\omega$ values  from  Eq.~(\ref{eq_ellipse}).  The  number of  signal events, $N_\omega$,
fitted and plotted as a function of the binned variable value, corresponds to the background-subtracted  distribution.

\begin{figure}[H]
\begin{minipage}{0.45\textwidth}
\centering
\includegraphics[width=1.1\linewidth
]{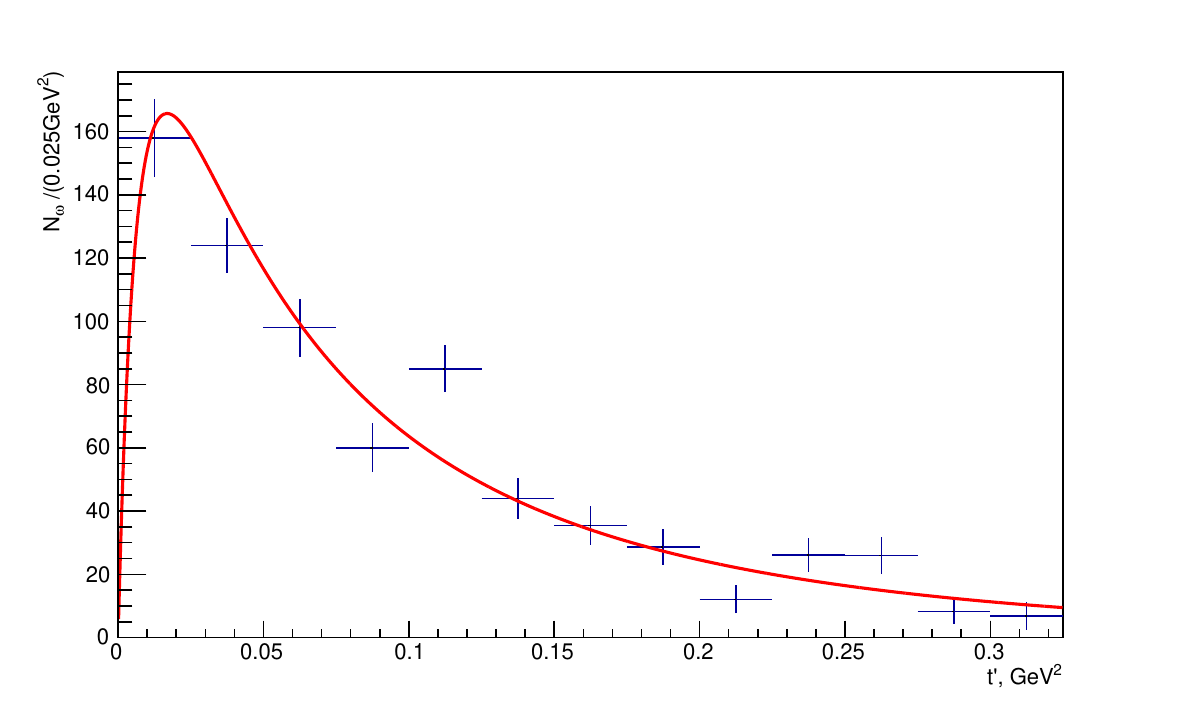}
\caption{$t'$ spectrum with non-$\omega$ background subtracted is presented (histogram with  error bars). The solid curve represents the fit with the OPE  model in Eq.~(\ref{ope}).}
\label{fig_filtered_tprime}
\end{minipage}
\hfill
\begin{minipage}{0.5\textwidth}
\centering
\includegraphics[width=1.\textwidth]{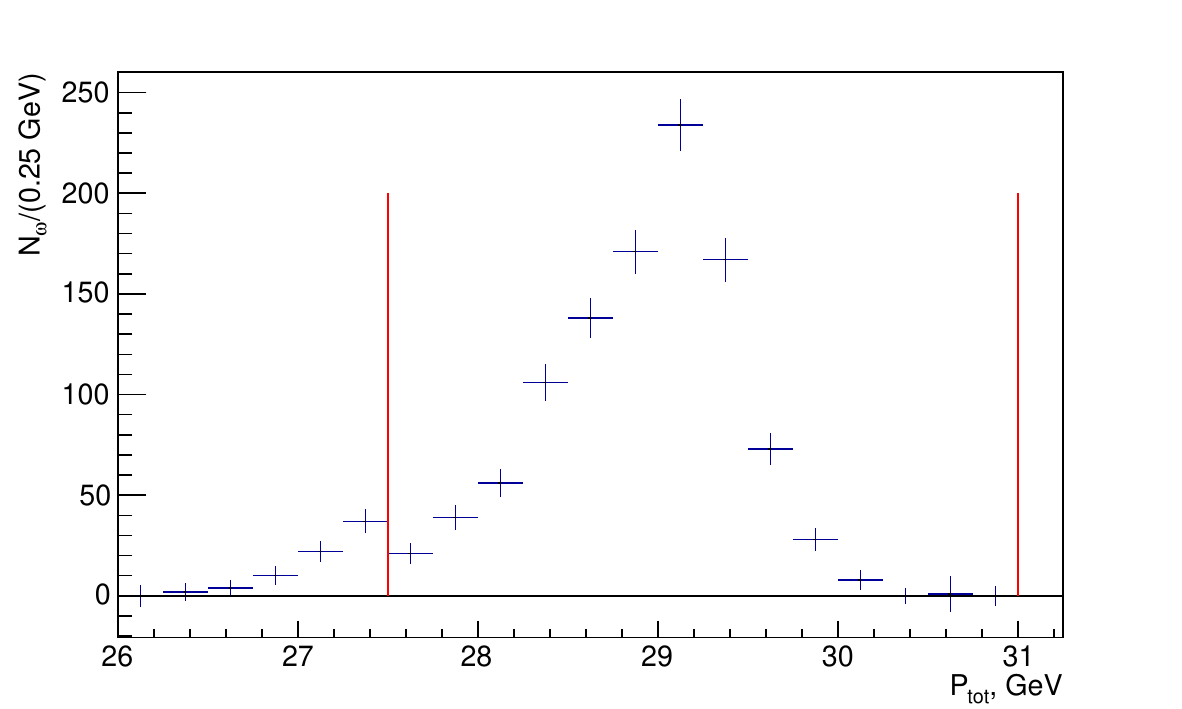}
\caption{$P_\text{tot}$ spectrum with the non-$\omega$ background subtracted. The red lines indicate   the applied exclusivity cut.}
\label{fig_filtered_ptot}
\end{minipage}
\end{figure}

The background-subtracted $t'$ spectrum is shown in Fig.~\ref{fig_filtered_tprime}, where 
$t' = \lvert t \rvert - \lvert t \rvert _\text{min}$.
\footnote{Here, $\lvert t \rvert _\text{min}$ is the minimum value  of 
$\lvert t \rvert $.} 
Since  $\lvert t \rvert _\text{min}$ is small,  
the difference between  $t'$ and $\lvert t \rvert$ can be ignored  in our kinematics. We will  use $t$ in the following sections. 
The momentum transfer is calculated  using the momenta of the beam  and the reaction products in the forward region, so that  nuclear effects including   Fermi motion do not influence this measurement. 
The uncertainty in $\lvert t \rvert$, determined by a  typical resolution of 8 MeV on the transverse momentum with respect to the beam, 
is $\sigma_{\lvert t \rvert} \approx 0.006$  GeV$^2$ for $\lvert t \rvert = 0.15$ GeV$^2$.

The neutral $\omega\phi$ system under study has  positive  $G$-parity and can be produced via one-pion exchange (OPE).
Fitting  the $t$ distribution with the shape
 \begin{equation}
\frac{dN_\omega}{dt} \propto \frac{t\, e^{\beta t}}{(t-m_\pi^2)^2}
\label{ope}
 \end{equation}
yields a  slope of $\beta=(4.3\pm 0.5) \, \text{GeV}^{-2}$.
However, the fit quality is poor  with $\chi^2/n.d.f = 36/11$. 

The slopes in the OPE parametrization,  obtained in Ref.~\cite{Hyams:1974wr}  for pion-induced production of 
 $\rho(770), f_2(1270) $, and $ \rho_3(1690)$ mesons at 17.2 GeV beam momentum,   range from 7.0  to  8.2 GeV$^{-2}$. 
The energy dependence of the slope for the $f_2$ can be found  in Ref.~\cite{f2_slope}. From this, we can evaluate the slope  for $q\bar{q}$ resonances  such as $\rho, f_2$, and $\rho_3$  to be approximately  $\beta \approx (8.2 \pm 0.2) \, \text{GeV}^{-2}$. This value    differs from the slope  we obtained for the $\omega\phi$ data.   The cause  is unknown and requires further  study.  
 
The  $P_\text{tot}$ spectrum after background subtraction, without  the exclusivity cut described in Eq.~(\ref{ptotcut}), is presented  in 
Fig.~\ref{fig_filtered_ptot} and was previously 
 discussed  in Sec.~\ref{sec3_ev_select}.

 The background subtraction of the distributions discussed  above was also performed by
fitting the $\phi$ peak in the $M_{2K}$ distribution while requiring $M_{3\pi}$ to be
within the $\omega$  band. 
The results obtained  are compatible with those  shown above. 

\section{Partial-Wave Analysis}\label{sec4_pwa}

A mass-independent partial-wave analysis of the $\omega\phi$ system was performed in 60 MeV wide $M_{\omega\phi}$ bins 
 ranging  from 1.78 to 2.80 GeV. 
 The $\lvert t \rvert$ range in  the PWA was chosen from 0 to  0.15 GeV$^2$ (see also below) without binning, resulting in  $N=580$ events. The aim is to suppress possible    contributions 
from production mechanisms other than OPE that have a 
wider $t$ distribution.    

In the  PWA model, the decay of the produced intermediate states to  $\omega\phi$ is approximated as a two-body decay.  
Symmetrization with respect to the final-state particles is not required, which distinguishes this analysis   from the otherwise similar PWA of the 
$\omega\omega$ system in Ref.~\cite{bib_om_om}.
  
 We use the extended maximum-likelihood method with the log-likelihood function 
\begin{equation} 
 \log\mathcal{L} = \sum_{k=1}^N \log  \sum_{\eta=\pm 1} \left\vert \sum_{i} T_i^{\eta} A_i^{\eta}(\tau_k)\right\vert^2 -  \sum_{\eta=\pm 1}   \sum_{i,j}  T_i ^{\eta} T_j ^{\eta *}\int  A_i^{\eta}(\tau) A_j^{\eta *}(\tau) \zeta(\tau) d \tau \quad ,
\label{log_lik}
\end{equation}
which is defined such that  the indices $i$ and $j$ enumerate  the partial-wave amplitudes. For each value of $i$, the total amplitude is  factorized into the unknown  production amplitude $T_i$ and the known decay amplitude $A_i$.   
To account for parity conservation at the  $\omega\phi$ production
vertex, the reflectivity basis  \cite{Chung} is used. In the high-energy limit, the
reflectivity quantum numbers  $\eta= +1$ and $\eta= -1$ correspond  to  natural and unnatural  parity  of the   exchange particle, NPE and UPE, respectively (Refs.~\cite{GJ,NC}).
The $T_i^{\eta}$ are estimated by maximizing Eq.~\ref{log_lik}   for each $M_{\omega\phi}$ bin independently.

The  custom PWA program calculates the first and second derivatives of $\log\mathcal{L}$ with respect to the fit  parameters $T_i^{\eta}$ 
analytically, resulting in faster fit convergence.

The decay amplitudes $A_i^{\eta}(\tau_k)$ are calculated for each event $k$ with   coordinates $\tau_k$ in the phase 
space. The construction of the decay amplitudes is based on Zemach's  non-relativistic tensor  formalism, as described   in Ref.~\cite{Zemach1}, which  we extended  to cover the case of two particles with non-zero spins.
 In our case, we  first construct two tensors of rank $j=1$:  one from the $K^+$ three-momentum vector $\vec{q}_1$   
in the $\phi$ c.m.s,  and one from the normal $\vec q_2$ of the  $\omega$ decay plane,  which is given by  the direction of  $\vec {p}_{\pi^-}\times\vec{p}_{\pi^+}$ in the $\omega$ rest frame. 
Using  these tensors, a rank-$S$ tensor is constructed, where $S$ is the total 
intrinsic spin of the $\omega\phi$ system. For $S = 2$, this tensor is symmetric and traceless. Next, we  construct a rank-$L$ tensor using  the three-momentum of the $\omega$ in the $\omega\phi$
rest frame. Here,  $L$ represents the orbital angular momentum between $\omega$ and $\phi$. Finally, we couple the $L$ and $S$ tensors  to a tensor of rank $J$, where $J$ is the spin of the intermediate state.  This tensor is further projected according to the spin-projection quantum number
$0\leq M \leq J$ with respect to the beam direction in the Gottfried-Jackson frame of the $\omega \phi$ system. The wave notation is 
 $J^{PC}M^\eta LS$. For instance, the decay amplitude for the wave $0^{++}0^-00$ has the structure 
$A \propto (\vec q_1 \cdot \vec q_2)$.

To account for the line shapes of $\omega$ and $\phi$ the decay amplitudes include square roots of 
corresponding Gaussian functions in $M_{3\pi}$ and $M_{2K}$, respectively. This helps to separate the non-$\omega$ 
and/or non-$\phi$ backgrounds from the signal process.
To account for these backgrounds and  
imperfections of the PWA model, a so-called FLAT amplitude is added incoherently to the  model in Eq.~(\ref{log_lik}). 
This amplitude uniformly fills  the $KK\pi\pi\pi$ phase space, i.e. $A_\text{FLAT} = \text{const}$. 
 
The matrix of the  normalisation  integrals in the second term of  Eq.~(\ref{log_lik}) contains 
the acceptance function $\zeta(\tau)$. The decay amplitudes are normalised such  that $\int \lvert A_i^\eta \rvert ^2 d\tau =1$. 
This gives   acceptance-corrected wave intensities $I_i^{\eta} = \left\vert T_i^{\eta} \right\vert^2$,  
expressed in terms of the number of produced events, with relative phases 
$\phi_i^{\eta} - \phi_j^{\eta} = \text{arg}(T_i^{\eta} T_j ^{\eta *} )$.
The integrals are pre-calculated once before the fit procedure  using  the Monte Carlo (MC) technique. The MC 
simulation of the experimental  setup and of the passage of particles through this setup  
is based on the Geant4 10.5 package 
\cite{GEANT4:2002zbu}. The reconstruction and selection procedures  applied to the simulated data are the same as for the real data.
 The 
 performance of the experiment was generally the same over  the four running periods used in this analysis, with minor differences in details, such as changes in the  beam momentum within 
0.5 GeV and in the position of detectors within 1 cm and others.  Separate MC models were used for all 
significant time periods.

It was checked that both the general characteristics of the event samples and the results of the PWA for the different running periods were  
compatible. Then the PWA was performed on the combined data sample. For this PWA, we used the total MC 
sample, where the number of  MC events generated for each  run period  is
proportional to the number of  
events recorded in the  period  divided by the corresponding efficiency. This means  that 
we average the efficiency over the run periods 
with weights proportional to the number of  events produced in each run period.

\begin{figure}[H]
    \centering
    \includegraphics[width=1.\linewidth]
    {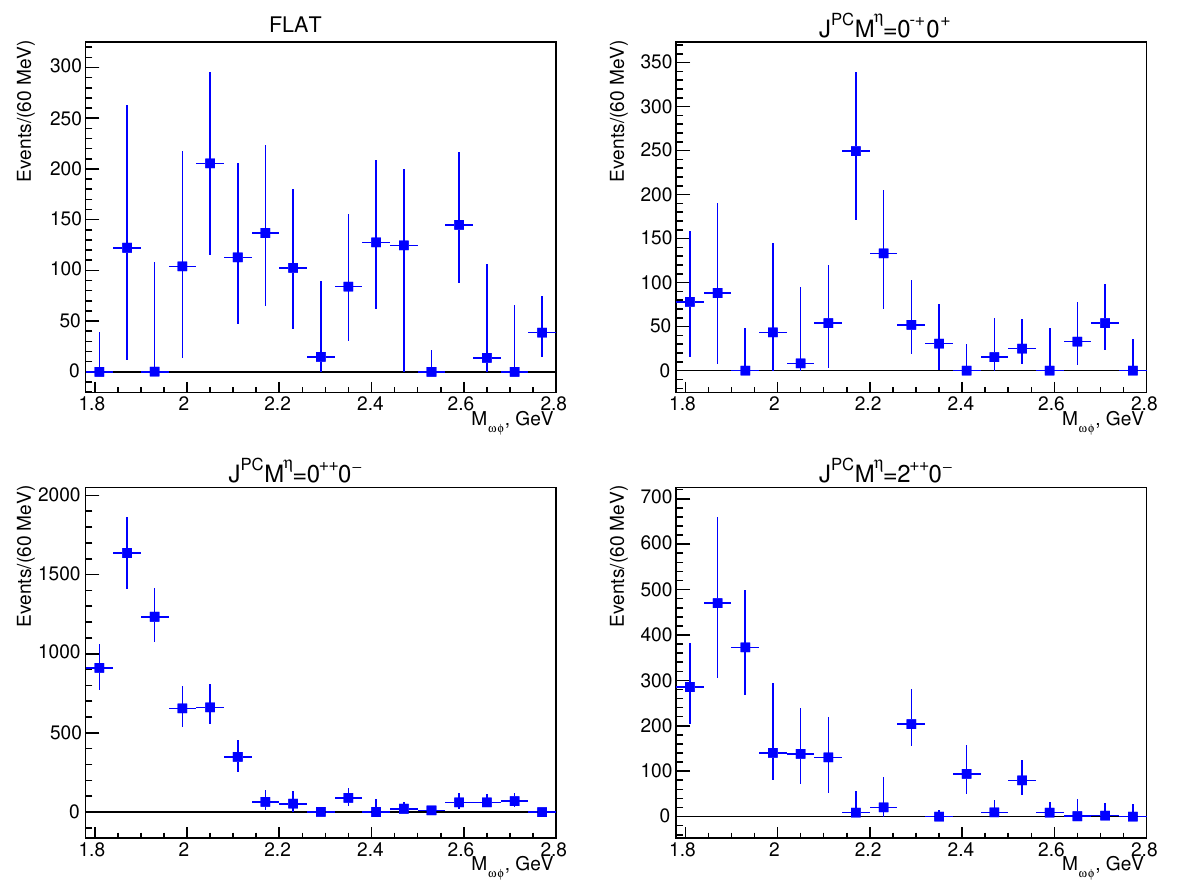}
    \caption{Partial-wave intensities obtained from  the final  PWA result  with $\lvert t \rvert <0.15\,\text{GeV}^2$. 
 }
    \label{fig_compare_waves_tprime}
\end{figure}

A small set of waves proved sufficient to describe  the data well. These waves are: FLAT, $0^{++}0^- 00$, $2^{++}0^- 02$,
$0^{-+}0^+ 11$. 
The intensity distributions of these
waves obtained from the PWA fit are shown  in Fig.~\ref{fig_compare_waves_tprime}. None of the waves can be excluded from the PWA fit without significantly degrading  the  quality of the fit. Other waves, when added, prove to be insignificant and  do not improve the fit quality. 
This justifies our choice of the minimalist PWA model. 

The intensities of the  waves integrated over $M_{\omega\phi}$  relative to the integral of the total intensity are  58\% for  $0^{++}$, 20\% for  $2^{++}$, 9\% for  $0^{-+}$, and 13\% for  FLAT.
The scalar wave with $J^{PC}=0^{++}$ is the dominant wave and has  a peak at threshold.
The  tensor wave $2^{++}$ also shows  some peak behaviour but it is much less pronounced.
The other two waves are smaller and have no  significant structures.

To check the stability of the analysis with respect to the chosen  cut value $\alpha_0$ for the PID likelihood ratio (see Sec.~\ref{sec3_ev_select}), 
the PWA was re-run with    $\alpha_0=1$ and $\alpha_0=4.$
The resulting partial-wave amplitudes remain stable within statistical uncertainties. In the absence of a MC model for the 
misidentified background, we take  this as  evidence for the smallness of this background.

To reveal the role of the non-$\omega\phi$ background, we performed the PWA with a looser elliptical cut in the ($M_{2K}$, $M_{3\pi}$) plane, i.e. using Eq.~(\ref{eq_ellipse})
with $\sigma_\phi = 8.8$ MeV and $\sigma_{\omega} = 22$ MeV. The only statistically significant 
change in the wave intensities is an increase of the FLAT wave, which is roughly proportional to  the increase in elliptical   
area. This suggests that the non-$\omega\phi$ background dominates in the FLAT wave.

The UPE waves make up 78\%  of the total intensity of the process  under study,   which 
supports the hypothesis of the OPE production mechanism. 
An alternative PWA with a wider  cut of $\lvert t \rvert <0.5 \,\text{GeV}^2$ 
is expected to have  a smaller  contribution from the UPE waves. The fit results in  relative intensities of  56\% for  $0^{++}$, 17\% for  $2^{++}$, 8\% for  $0^{-+}$, and 20\% for  FLAT waves. This indicates a decrease in the contribution of the UPE waves to 73\%.   

\section{Results and Discussion}\label{sec6_results}

The wave intensities in the final PWA results for $\lvert t \rvert < 0.15 \,\text{GeV}^2$ are presented in Fig.~\ref{fig_compare_waves_tprime}.
Due to a positivity constraint and small sample size,  evaluating  errors of  PWA results  based on the 
inverse of the Hessian matrix of the
log-likelihood function at its maximum
can be biased. Instead, we evaluate intensity errors, represented as asymmetric bars,  by   scanning  the log-likelihood value as a function of the intensity near its  maximum
and determining the intensity interval that
corresponds to a decrease of the maximum log-likelihood by half a
unit.
 
\begin{figure}[ht]
    \centering
    \includegraphics[width=0.6\textwidth]{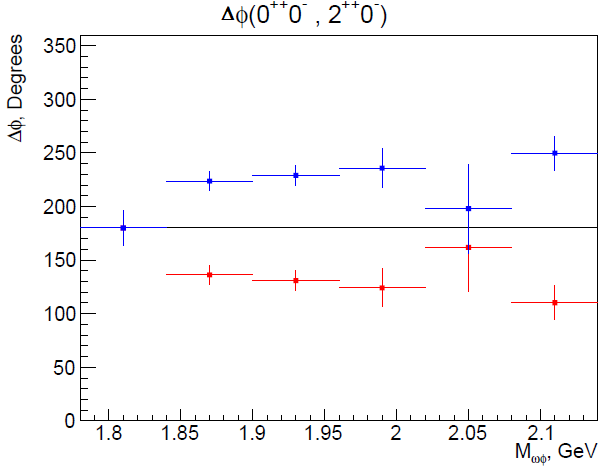}
    \caption{Phase difference between the   $0^{++}$  and the  $2^{++}$ waves. Two solutions are shown with different colours. The horizontal line indicates 180 degrees. }
    \label{phase}
\end{figure}

In this PWA, only the phase between the   $0^{++}$  and the  $2^{++}$ waves is available. However, in the  case of narrow resonances the measurement of the  phase is twofold ambiguous. Two solutions are shown in the Fig.~\ref{phase} for a limited mass range.  For  higher masses,   the phase cannot be  measured  due to the smallness of the intensities.  There is no definitive  interpretation for the phase motion in the  $0^+$ wave due to  the ambiguity and  the absence  of a model for the reference  $2^+$ wave.  

In the range of $M_{\omega\phi}$ shown, the total  efficiency, including the cuts,  varies slowly with the $M_{\omega\phi}$  between 5.1\% and 6.7\%.    
A  comparison of the measured  distributions with those obtained from  the Monte Carlo events weighted with the PWA result is shown  in Figs.~\ref{m2g},\ref{fig_predict_m_KK},\ref{fig_predict_m_pipipi0},\ref{fig_lambda},\ref{cosGJ_om}  and
\ref{fig_angle_om_kaon}. The model describes the experimental data rather well.   
  The Monte Carlo events were  generated  using Breit-Wigner amplitudes for  $\omega$ and $\phi$, with their nominal masses and widths.

\subsection{Comparison of  $\omega\phi$ and $\omega\omega$ channels}

In this section, we compare the characteristics of the reactions $\pi^- \text{Be} \to \text{A} \omega\phi$ and 
$\pi^- \text{Be} \to \text{A} \omega\omega$  with the predictions of the OZI rule \cite{OZI1,OZI2,OZI3}. The  characteristics of these channels in charge-exchange reactions and in radiative $J/\psi$ decays are also compared,  and it is found that the same signal is observed.

   To this end, we use  the results of a PWA of the $\omega\omega$ system performed on data obtained with the VES setup
at 28 GeV nominal beam momentum  \cite{bib_om_om, VES:2010uah}.
In the following  text, values from the $\omega\omega$ analysis are denoted by "28", 
while the values from our current  analysis are denoted by  ”29”.
Here, these numbers indicate the slightly different beam momenta for
the two data samples.

The data for $\omega \omega$  are dominated  by $J^{PC} = 2^{++}$ waves across  a broad mass  range. 
 The intensity $N^{0^+}_{\omega \omega, 28}$ of the smaller $0^{++}0^-00$ $\omega\omega$ wave was extracted from  Fig.~4 of Ref.~\cite{VES:2010uah}, while  the total 
intensity $N_{\omega \omega, 28}$
was obtained by summing the UPE and NPE wave intensities 
(dashed histograms  in Fig.~4 of Ref.~\cite{bib_om_om}).
The intensity of the $0^{++}$ wave is 
significant but  lacks  prominent structures. 
From these intensity distributions, we derive the relative contribution of the scalar wave, $N_{\omega\omega,28}^{0^+}/ N_{\omega\omega,28}$,  
as a function of $M_{\omega\omega}$.

\begin{figure}[H]
\centering
\includegraphics[width=0.6\linewidth]
{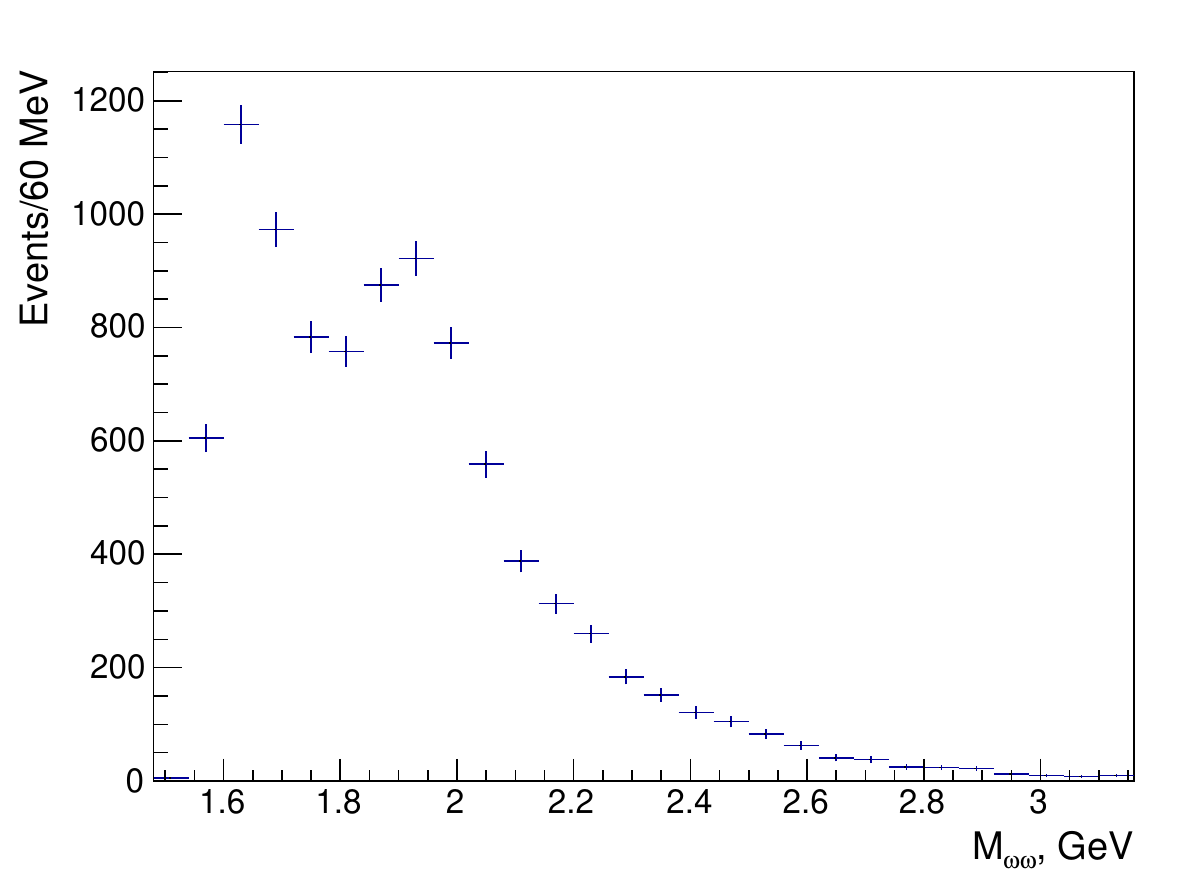}
\caption{The $\omega\omega$ invariant mass spectrum for the 29-GeV data sample.}
\label{fig_mass_om_om_new}
\end{figure}

To establish a relationship between  the 28-GeV and 29-GeV data samples,  we need to know    
  the $M_{\omega\omega}$ spectrum  $N_{\omega\omega,29}^{\text{obs}}$  from the latter one. 
The   selection requirements applied and the values for the cuts are in general similar  to the  $\omega\phi$ case. Characteristics   specific  to $\omega \omega$ are listed below.  Particle identification is not applied to the four charged particles.  Four to five photon clusters in the EMC are required. Two photons with an invariant mass within 25 MeV of the nominal   $\pi^0$ mass  are considered as neutral pions candidates. 
We construct all disjoint combinations of pairs of  $\pi^+\pi^-\pi^0$ subsystems, where we allow
for one unassigned photon cluster. We select the pair with the
smallest distance from the point that corresponds to nominal $\omega$
masses in the $(M_{3\pi},M_{3\pi})$  plane within a circular region of 30 MeV
radius about that point. Figure~\ref{fig_mass_om_om_new} shows the $M_{\omega\omega}$ spectrum after all
selection cuts.

Next, the $N^\text{obs}_{\omega \omega, 29}$ spectrum is corrected by the efficiency 
$\epsilon_{29}(M_{\omega\omega})$. It is estimated from a Monte Carlo simulation with the full chain of reconstruction and selection procedures. The event kinematics for the Monte Carlo  is 
generated using the measured  $t$ distribution and the known distribution of  $\lambda$ for the two 
$\omega$ decays.  The simulation of angular distributions is based on the dominant wave with  $J=2$ and  $L=0$.  The value of  $\epsilon_{29}$ is found to be  weakly dependent  on  $M_{\omega\omega}$ in the analysed  range, with an average value of $\left<\epsilon_{29}\right> \approx 0.06$. To estimate the systematic uncertainty of the efficiency, Monte Carlo data were also generated  for the  $J=0 \, ,L=0$  wave. 
The efficiencies estimated using these data differ by typically 5\%
from the ones used in the main analysis.

With the above, the scalar wave intensity  for the 29-GeV data is:
\begin{equation}
N_{\omega\omega,29}^{0^+}(M_{\omega\omega})=\frac{N_{\omega\omega,29}^{obs}}{\epsilon_{29}}  \frac{N_{\omega\omega,28}^{0^+}}{N_{\omega\omega,28}} \, C_t\quad .
\label{eq_intensity_om_om_new}
\end{equation} 
To account for the wider  range of $\lvert t \rvert < 0.20\,\text{GeV}^2$ used in the analysis of the  28-GeV  data, an additional 
correction factor of $C_t = 1.13 \pm 0.03$ is applied. 

\begin{figure}[H]
\begin{minipage}{0.45\textwidth}
\centering
\includegraphics[width=0.9\linewidth]
{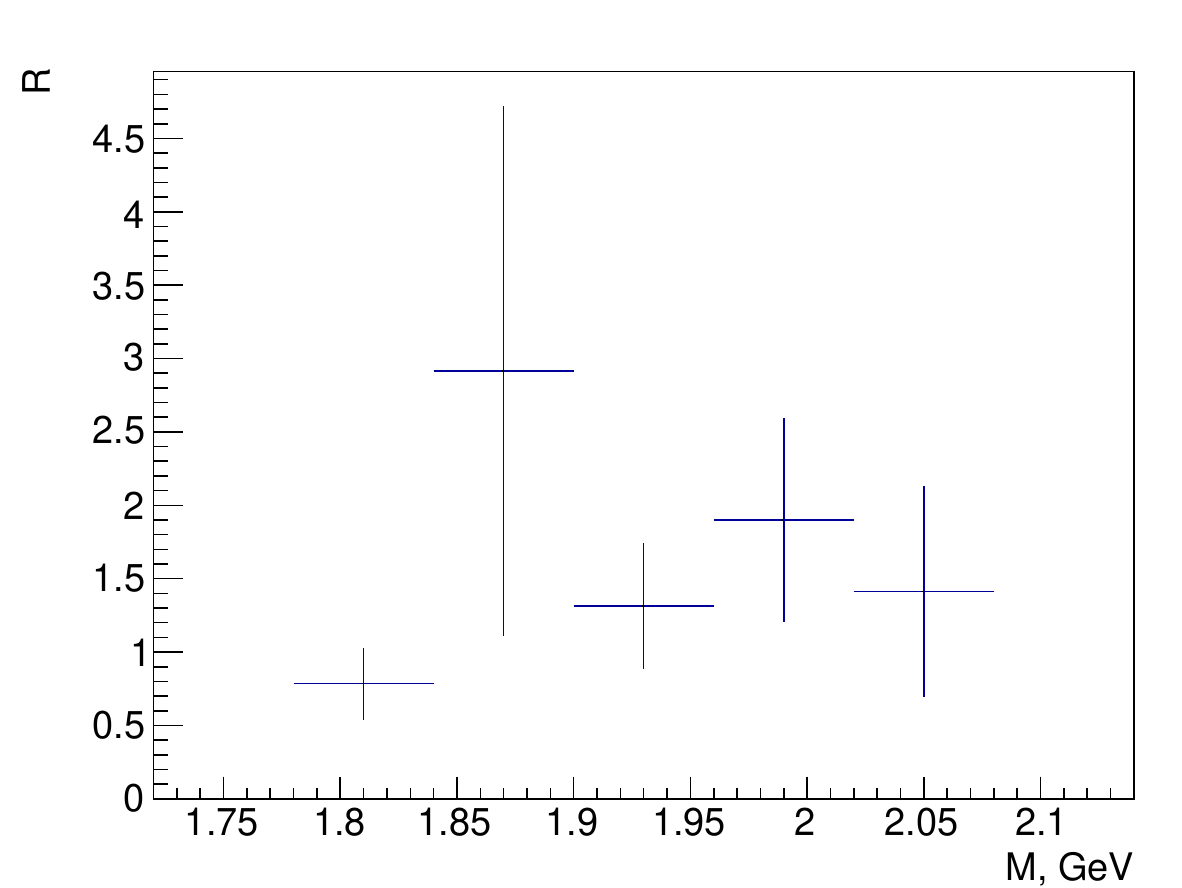}
\caption{Ratio of the $J^{PC}=0^{++}$ wave intensities in the $\omega\phi$ and $\omega\omega$ channels  as a function of mass (see Eq.~(\ref{eq_R})).
}
\label{fig_ratio_0plus}
\end{minipage}
\hfill
\begin{minipage}{0.45\textwidth}
\centering
\includegraphics[width=0.9\linewidth]
{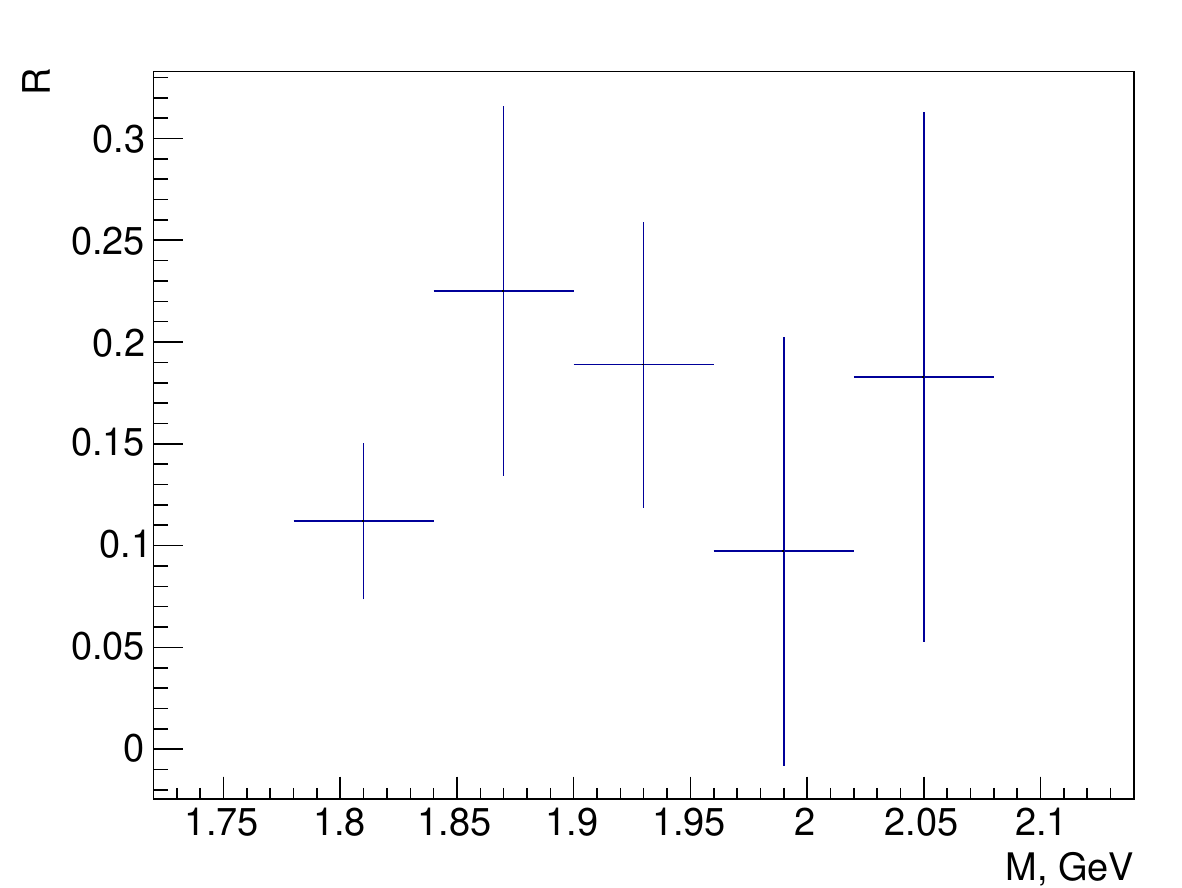}
\caption{Similar ratio as in Fig.~\ref{fig_ratio_0plus} but for the $J^{PC}=2^{++}$ wave. 
}
\label{fig_ratio_2plus}
\end{minipage}
\end{figure}

Finally,  in  the  mass region where both channels are open and   in the 
range $\lvert t \rvert < 0.15\,\text{GeV}^2$,  the ratio 
\begin{equation}
R =\frac{N_{\omega\phi,29}^{0^+}}{N_{\omega\omega,29}^{0^+}}  \frac{Br(\omega \rightarrow \pi^+\pi^-\pi^0)}{Br(\phi \rightarrow K^+K^-)} \quad 
\label{eq_R}
\end{equation}
of the $0^{++}0^-{00}$ wave intensities in the $\omega\phi$ and $\omega\omega$ decay channels is calculated. 
As shown  in Fig.~\ref{fig_ratio_0plus}, $R$ is close to 1.
 The error bars  encompass   the statistical and systematic uncertainties of 
all factors in Eqs.~(\ref{eq_intensity_om_om_new}) and (\ref{eq_R}). The partial-wave intensities give the largest contribution.

We also calculate  the ratio 
\begin{equation}
R_A = R \frac{q_{\omega\omega}}{q_{\omega\phi}}
\label{ra}
\end{equation}
which is determined by the squared amplitudes  of the two processes. Here, we account for the different phase-space by dividing the acceptance-corrected intensities  by the corresponding 
breakup momenta  $q_{\omega\omega}$ and $q_{\omega\phi}$ that are defined in the c.m.s. of the $\omega\omega$ and  $\omega\phi$ system, respectively.  
The lowest  mass bin 
is omitted due to the very rapid change of the breakup momentum near the threshold.  Averaging  the four bins in the mass 
range from  1.84 to 2.08 GeV  yields a value  of $\left<R_A\right> (0^{++}) = 2.4\pm 0.5$ with  $\chi^2/\text{ndf} = 1.5/3$.  

Applying the same procedure to the $2^{++}0^-02$  wave, we obtain  $R$ values, which are shown  in  Fig.~\ref{fig_ratio_2plus}. They results in $\left<R_A\right> (2^{++})= 0.30 \pm 0.08$ with $\chi^2/\text{ndf} = 2.3/3$. 

The obtained $\left<R_A\right>$ values can be compared to predictions based on the OZI rule \cite{OZI1,OZI2,OZI3}. For  the interaction of particles composed of  light quarks it states that  the  relative  yield of the $\phi$ to  $\omega$ mesons is determined   by $\tan^2{\theta} = 4.2\cdot 10^{-3}$. Here the  mixing angle $\theta = 3.7^\circ$ \cite{nomokon} is defined through the $\phi$ quark content: $\phi=s\bar s \cos{\theta}+n\bar n\sin{\theta}$. 
Experimental data on various reactions \cite{nomokon} indicate a significant suppression of reactions involving $\phi$ production, although this suppression is sometimes far  less than the theoretical value.  

The  $\left<R_A\right>(0^{++})$ value observed in our experiment did not  show any  suppression and is larger than that in all other  experiments. The cause of this significant OZI rule violation is currently unknown. 
One possible explanation for this phenomenon is the substantial mixing of scalar states with different quark compositions. This mixing has been observed experimentally and can be explained by theoretical models \cite{to_OZI}. In contrast, the mixing for the tensor states is much less \cite{to_OZI}, resulting in  $\left<R_A\right>(2^{++})$ value that is  closer to the OZI prediction. 

The structure we observe in the scalar $\omega\phi$  wave is similar to  the $X(1810)$ reported in  radiative $J/\psi$ decays to  $\omega\phi$ in Refs.~\cite{BES:2006vdb,BESIII:2012rtd}.  These results are included  by the PDG  in the entry for the $f_0(1710)$  \cite{PDG}. 
In order to verify that one and the same object is observed in the charge-exchange reaction and in the radiative $J/\psi$ decays, we now compare the relative intensities of the production of the $J^{PC}=0^{++}$ wave in the $\omega\omega$ and $\omega\phi$ channels  in  these two reactions. 

From the PDG entry for the  $J/\psi$ we  get 
\begin{equation}
R_\text{Rad.Dec.} = \frac{Br(J/\psi\rightarrow \gamma f_0(1710) \rightarrow \gamma \omega\phi)}
{Br(J/\psi\rightarrow \gamma f_0(1710) \rightarrow \gamma \omega\omega)} = 0.8\pm 0.4 \quad .
\label{r_rad}
\end{equation}
The branching  fraction for the $\omega\omega$ channel  in  Eq.~(\ref{r_rad}) was measured for  masses starting  at  the  $\omega\omega$ threshold.   
 For comparison,   we now calculate the corresponding ratio for the charge-exchange reaction by summing  the numerator  and denominator in Eq.~(\ref{eq_R})  over the  mass bins from the $\omega\phi$ and  $\omega\omega$ thresholds, respectively, up to   2.08 GeV. The extended  mass  range for the $\omega\omega$ channel  leads to a lower  value of $R_\text{Ch.Ex.} = 0.70\pm 0.15$,  compared to   the  values shown in Fig.~\ref{fig_ratio_0plus}. This  value of $R_\text{Ch.Ex.}$ is consistent  with   $R_\text{Rad.Dec.}$.

Both the shape of the $0^{++}$ wave intensity in the $\omega\phi$ channel and its ratio to that in the $\omega\omega$ channel are 
consistent in  radiative $J/\psi$ decays and in  charge-exchange reactions. 
This  suggests  that the same object is observed in both   reactions.

\subsection{Parameters of Scalar Resonance
}

We  further consider two alternative assumptions regarding the   identification of the observed signal. 
In the first assumption,  the source of the signal is the $f_0(1710)$ from Ref.~\cite{PDG}   with parameters given  in Eq.~(\ref{f0_params}), 
 and in the second  it is the $f_0(1770)$ from Ref.~\cite{Sarantsev:2021ein} with  parameters given in Eq.~(\ref{f01770_params}). 
 
Due to the proximity  of the  $f_0(1710)$ mass to the $\omega\phi$ threshold,  we describe  
the wave intensity   with a  Flatt\'e parameterisation, i.e.  
\begin{equation}
\frac{dN}{dM} = 
\frac{C M_R^2 \Gamma_0 g q_{\omega\phi}}{(M_R^2-M^2)^2+M_R^2(\Gamma_0+gq_{\omega\phi}) ^2} \quad\quad (M>M_\omega+M_\phi) \qquad .
\label{Flatte}
\end{equation}
Here, $\Gamma_0$ is a constant partial width that accounts for the decay into   channels that  are far above their thresholds, and $gq_{\omega\phi}$  accounts for the  mass-dependent partial width for  the  $\omega\phi$ channel. 
The shape of the curve is determined by three parameters: $M_R\,,\Gamma_0$, and $g$, which  are strongly correlated in the fit. 
However, when using Eq.~\ref{Flatte} for numerical calculations, the branching fraction into the
 $ \omega\phi $   channel closely approximates the value of $g$. This holds true regardless of
the two other parameters that were reasonably chosen.
To better constrain the model and stabilize the fit, we set the value of $g$  to a  middle value within  the range of possible branching fractions.     
An upper limit for this range is conservatively  estimated  using   data from Ref.~\cite{PDG}. Specifically,  
\begin{equation}
    Br_\text{max}(f_0(1710)\rightarrow \omega\phi)  = \frac{\Gamma(J/\psi \rightarrow \gamma f_0(1710)\rightarrow \gamma \omega\phi)}{\Sigma \Gamma(J/\psi\rightarrow \gamma f_0(1710)\rightarrow \text{all observed channels})} = 0.14 \,.
    \label{br_f0}
\end{equation}
The minimum value of $Br_\text{min}(f_0(1710)\rightarrow \omega\phi)$ is 0.05,   as determined later  in  Eq.~(\ref{jpsi_to_f0}).

\begin{figure}[H]
\centering
\includegraphics[width=0.7\textwidth,trim={0 0.3cm 0 0.7cm},clip]
{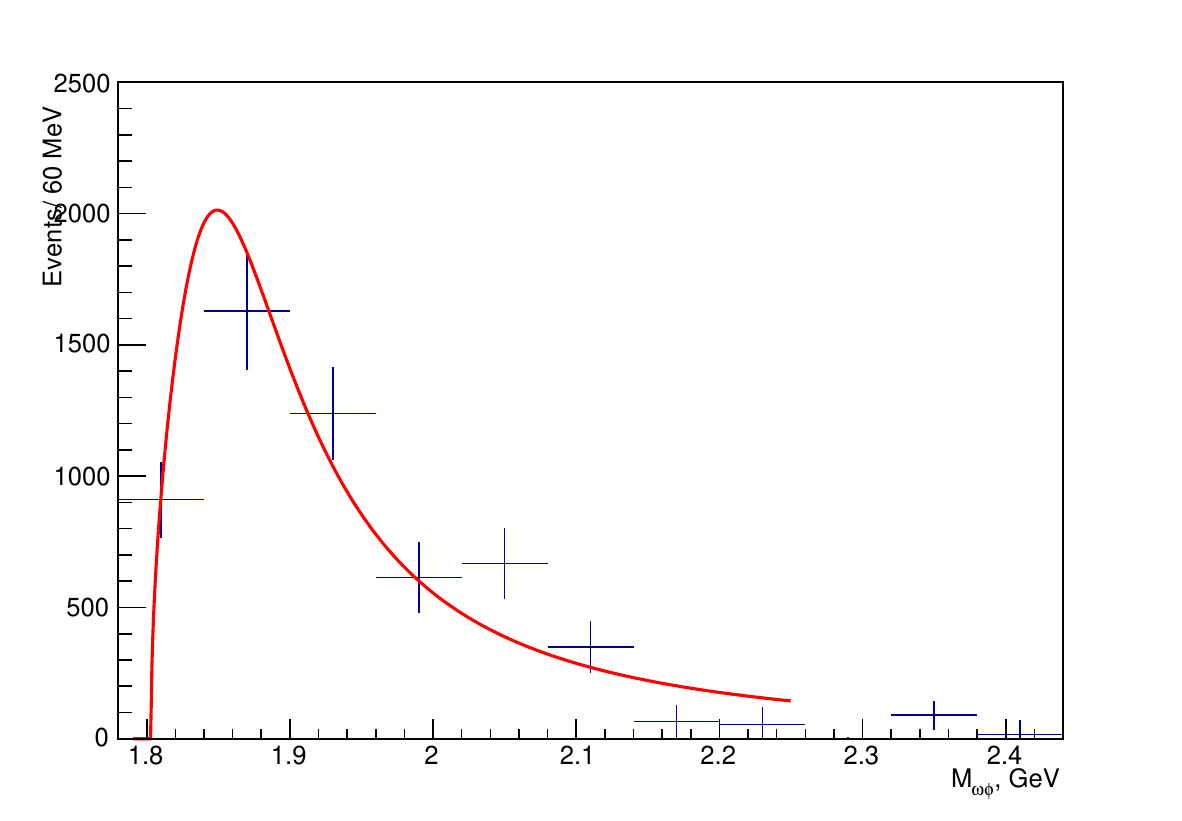}
\caption{ The intensity of wave $0^{++}0^-00$   in the $\omega\phi$ channel (points from Fig.~\ref{fig_compare_waves_tprime}) fitted with the model function in Eq.~(\ref{Flatte}) with  $g=0.1$.}
\label{fig_BW_fit}
\end{figure}

Fitting Eq.~(\ref{Flatte}) to the $0^{++}$ wave intensity with a 
fixed value  $g=0.1$ yields   resonance parameters of $M_R=(1834 \pm 14)  \,\text{MeV}$ and  
$\Gamma_0 = (114 \pm 15 )  \,\text{MeV}$, with a $\chi^2_\text{min}/\text{ndf}=13.3/5$  (see  Fig.~\ref{fig_BW_fit}). 
Systematic uncertainties of the fit parameters were  estimated by varying   $g=0.10\pm 0.05$.  
So the  resonance parameters of Eq.~(\ref{Flatte}) are 
$M_R=(1834 \pm 14 \,\text{(stat.)} ^{+2}_{-10} \,\text{(syst.)}) \,\text{MeV}$ and  
$\Gamma_0 = (114 \pm 15 \,\text{(stat.)} ^{+5} _{-15} \,\text{(syst.)}) \,\text{MeV}$. 
 We verified  that  altering the range of the fitted mass  does  not affect  the outcome. 
   
The mass and width   are compatible  with the parameters  of the $X(1810)$ reported in Ref.~\cite{BESIII:2012rtd} for $J/\psi \to \gamma \omega \phi$ decay: $M=(1795 \pm7 ^{+23}_{-20})$ MeV and $\Gamma =(95\pm 10 ^{+78}_{-82})$ MeV .  

\subsection{Branching Fractions of Scalar Resonance}

The intensity of the scalar wave integrated over the mass region $M_{\omega\phi} <
2.14$ GeV, where it is significant,   amounts to  $5.4\cdot 10^3$ events. 
Accounting for  the branching fractions for the $\omega$ and $\phi$ decays  from PDG~\cite{PDG}, this corresponds to a  
cross section of
$$\sigma(\pi^- \text{Be} \rightarrow  \text{A}\,  \omega\phi) \, (J^{PC}=0^{++},\, M_{\omega\phi}<2.14\,\text{GeV}, \lvert t \rvert <0.15\,\text{GeV}^2 )  = $$
\begin{equation}
98 \pm 7 \text{(stat.)} \pm 7\text{(syst.)} \,\text{nb}  \quad .
\label{sigma_pibe}
\end{equation}
The  systematic uncertainty is  calculated  as the quadratic sum of  the uncertainties in the efficiency,
 cross-section normalization, and the two branching fractions.  

We will use the  cross section in Eq.~(\ref{sigma_pibe}) to evaluate  the value of $Br(f_0\rightarrow\pi\pi)Br(f_0\rightarrow\omega\phi)$ using the one-pion-exchange model. To do this,  
we need to know the cross section for the charge-exchange reaction 
on the proton. Several  theoretical and experimental works have related 
the cross sections of charge-exchange reactions on the proton to those on nuclei \cite{Kolbig:1968rm,Guisan:1971up,Apokin:1988sj}.
For light nuclei, the $Z$ dependence of the cross section
 can be approximated by $\sigma \propto  Z^{\alpha}$, where $Z$ is the  number of protons. Based on  experimental data 
on the productions of  isoscalars  in charge-exchange reactions at a  beam momentum of 39 GeV \cite{Apokin:1988sj}, we 
estimate that  $\alpha = 0.73\pm 0.03$. For Beryllium,  this corresponds to  $Z^\alpha = 2.7\pm 0.2$. Using  Eq.~(\ref{sigma_pibe}), we obtain       
$$
\sigma(\pi^- p \rightarrow n\, \omega\phi) \,(J^{PC}=0^{++}, \, M_{\omega\phi}<2.14\,\text{GeV}, \, \lvert t \rvert <0.15\,\text{GeV}^2)  = 
$$
\begin{equation}
(36. \pm 5.) \, \text{nb} \quad .
\label{sigma}
\end{equation}
The systematic uncertainty dominates the  total uncertainty.
Here and below  a PDG rounding rule is applied to  numerical results.

We  then proceed to  calculate  the  product of  $Br(f_0(1710)\rightarrow \pi\pi)$ and $Br(f_0(1710)\rightarrow \omega\phi)$
using  the OPE approximation for the  production of a resonance $X$  decaying to a given channel 
\cite{Chew:1958wd,Williams:1970rg}:
\begin{displaymath}
\frac{d\sigma(\pi^-p\rightarrow n X)  Br(X\rightarrow \text{channel})}{d \lvert t \rvert} = {}
\end{displaymath}
\begin{equation}
26.9 \,\text{mb} \, Br(X\rightarrow\pi\pi)  Br(X\rightarrow \text{channel})  \frac{ M_X\Gamma_X} {P_\text{beam}^2}  \frac{\lvert t \rvert \, e^{\beta (t-m_\pi^2)}}{(t-m_\pi^2)^2}  \quad .
\label{ope_approx}
\end{equation} 
This approximation is commonly  used to analyse  charge-exchange reactions with  UPE dominance. The systematic 
uncertainty is estimated to be approximately  20\% \cite{Hyams:1974wr}.

Integrating Eq.~(\ref{ope_approx}) over $\lvert t \rvert$ with $\lvert t\rvert _\text{max} = 0.15\ \text{GeV}^2$,  $\beta=(4.3 \pm 0.5) \,\text{GeV}^{-2}$ from Eq.~(\ref{ope}),  and $M_X \Gamma_X = (0.260\pm 0.021)\,\text{GeV}^2$ from Eq.~(\ref{f0_params}), and using  Eq.~(\ref{sigma}), we get  
\begin{equation}
Br(f_0(1710)\rightarrow \pi\pi) Br(f_0(1710)\rightarrow \omega\phi)= (4.8 \pm 1.2)\cdot 10^{-3} \quad .    
\label{br_product}
\end{equation}
 
Using Ref.~\cite{PDG}, we calculate the product     
\begin{equation}    
Br(J/\psi\rightarrow\gamma f_0(1710) \rightarrow \gamma\pi\pi) Br(J/\psi\rightarrow\gamma f_0(1710) \rightarrow \gamma \omega\phi) = (9.5\pm 2.6)\cdot 10^{-8} \quad .
\end{equation}
Using this value and Eq.~(\ref{br_product}),    we find  
\begin{equation}
Br(J/\psi\rightarrow\gamma f_0(1710)) = (4.5\pm 0.8)\cdot 10^{-3} \quad.
\label{jpsi_to_f0}
\end{equation}

If compared  with the experimental value for  the decay of the $f_0(1710)$ into the five channels 
$\pi\pi\, ,KK\, , \eta\eta \, , \omega\omega$, and $\omega\phi$ listed by PDG~\cite{PDG}
\begin{equation}
    Br(J/\psi\rightarrow\gamma f_0(1710))  Br(f_0(1710)\rightarrow 5\,\text{channels}) = (2.13 \pm 0.18)\cdot 10^{-3}  \quad, 
\end{equation}
 Eq.~(\ref{jpsi_to_f0})  leaves a branching fraction for  unlisted  channels of
\begin{equation}
Br(f_0(1710)\rightarrow 4\pi, 6\pi, \eta\eta',\pi\pi KK,...) = (2.3\pm 0.8)\cdot 10^{-3} \quad.
\end{equation} 
This value does not contradict  the result of Ref.~\cite{4pi}, i.e. 
\begin{equation}
Br(f_0(1750) \rightarrow \sigma\sigma) = (9.0\pm 1.3) \cdot 10^{-4}  \,   , \,
Br(f_0(1750) \rightarrow \rho\rho) = (1.90\pm 0.14) \cdot 10^{-4}  \, ,
\end{equation}
assuming that $f_0(1710)$ and $f_0(1750)$ are  the same object. 

Comparing the   value  obtained for  $Br(J/\psi\rightarrow\gamma f_0(1710))$ in Eq.~(\ref{jpsi_to_f0}) with the value $(3.8\pm 0.9)\cdot 10^{-3}$ calculated 
for a scalar glueball in Ref.~\cite{Gui:2012gx} using quenched lattice QCD
suggests the presence of a significant or even dominant glueball  component in the $f_0(1710)$.  This branching fraction  is much lower for other known scalars \cite{glueball}.

To illustrate  the properties of the $f_0(1710)$, we refer  to the model presented in Ref.~\cite{Close:1996yc}. 
The model states that the branching of the radiative decay of a heavy vector quarkonium to a resonance $R_J$ with spin $J$ is proportional to the $R_J$ decay width into two gluons. This can be expressed as
\begin{equation}
\frac{Br(Q\bar Q_V \rightarrow\gamma R_J)}{Br(Q\bar Q_V\rightarrow \gamma gg)} = K_J(M_R/M_V)\frac{M_R\,\Gamma_R\,Br(R_J\rightarrow gg)}{M_V^2} \quad .
\end{equation}
The function $K_J(M_R/M_V)$  contains  a loop integral for   virtual gluons in the $Q\bar Q_V \rightarrow\gamma R_J$ decay. This integral  is calculated using  a model for the form factor of  $R_J$. 
Using  the value $K_0(M_{f_0(1710)}/M_{J/\psi})=0.116$  from  Ref.~\cite{Close:1996yc}, the branching fraction  $Br(J/\psi\rightarrow \gamma gg ) = 0.088\pm 0.011$ according to Ref.~\cite{PDG}
and the branching fraction  $Br(J/\psi\rightarrow\gamma f_0(1710))$    from Eq.~(\ref{jpsi_to_f0}), one can calculate the branching fraction  for the $f_0(1710)$ decay to two gluons:  
$Br(f_0(1710)\rightarrow gg) = 1.7 \pm 0.4$.  
The quoted uncertainty is primarily   due to the uncertainties of the $J/\psi$ radiative-decay branching fractions and the OPE model. It  does not include
the systematics  of the model from Ref.~\cite{Close:1996yc}.  
The measured branching fraction of about 1 
indicates a significant glueball component in the $f_0(1710)$.

 Ref.~\cite{Sarantsev:2021ein} provides  strong  evidence  that  the   $f_0(1710)$  reported in Ref.~\cite{PDG} is actually split into two states, an $f_0(1710)$, which is distinct from the $f_0(1710)$ in the PDG, and an $f_0(1770)$.
In the reaction $\pi^- p\rightarrow n\omega \phi$ we observe only one state, the $f_0(1770)$, while the other state, the $f_0(1710)$, does not contribute due to the smallness of its couplings to $\pi \pi$ and $\omega \phi$.
The resonance parameters  of the  $f_0(1770)$ reported in Ref.~\cite{Sarantsev:2021ein} are:  
\begin{itemize}
\item $M = (1765 \pm 15) \,\text{MeV}, \quad  
\Gamma = (180 \pm 20) \,\text{MeV}$ ,       
\item $Br(J/\psi\to\gamma f_0(1770)\to \gamma\pi\pi)=(2.4\pm 0.8)\cdot 10^{-3}$, 
\item $Br(J/\psi\to\gamma f_0(1770)\to \gamma\omega\phi)=(2.2\pm 0.4)\cdot 10^{-3}$.
\end{itemize}
Using these values in the above equations, we can determine the following values:  
\begin{itemize}
\item $Br(f_0(1770)\rightarrow \pi\pi) Br(f_0(1770)\rightarrow \omega\phi)= (3.9 \pm 1.0)\cdot 10^{-3}$, 
\item 
$Br(J/\psi\rightarrow\gamma f_0(1770)) = (3.7\pm 0.8)\cdot 10^{-3}$,
\item 
$Br(f_0(1770)\rightarrow 4\pi, 6\pi, \eta\eta',\pi\pi KK...) = (1.6\pm 0.9)\cdot 10^{-3}$,  
\item $Br(f_0(1770)\rightarrow gg)= 1.12\pm 0.32$.
\end{itemize}
These values do not affect  the conclusion regarding the significant presence of  a  glueball component in the studied   $f_0$ state.

\section{Conclusions}\label{sec_conclusion}
The reaction $\pi^- \text{Be} \rightarrow \text{A}\,\omega\phi$ was studied at a  beam momentum of 29 GeV. 
The $J^{PC}=0^{++}$ wave  dominates in the $\omega\phi$ system  and exhibits a threshold enhancement. 
The average ratio of the intensities
of the 
$0^{++}$ wave in the $\omega\phi$  and $\omega\omega$ channels, corrected for phase space, was found to be
$\left<R_A\right> =2.4\pm 0.5$. 
This indicates  a significant  violation of the OZI rule. 
 
We measure a cross section of  
$$
\sigma(\pi^- \text{Be} \rightarrow \text{A}\,  \omega\phi) \,(J^{PC}=0^{++}, \, M_{\omega\phi}<2.14 \,\text{GeV},\,\lvert t \rvert <0.15\,\text{GeV}^2 )  = 
$$
$$ 98 \pm 7 \text{(stat.)} \pm 7\text{(syst.)} \quad\text{nb} \,.$$
The  signal  in the  $\omega\phi$ channel  can be attributed to the known $f_0(1710)$ from Ref.~\cite{PDG} or to the $f_0(1770)$ from Ref.~\cite{Sarantsev:2021ein}. 
Using the one-pion exchange model for the reaction  $\pi^- p \rightarrow n\, f_0$ and the  branching fractions for the radiative
$J/\psi$ decays,  a branching fraction of
$Br(J/\psi\rightarrow\gamma f_0(1710)) = (4.5\pm 0.8)\cdot 10^{-3} \,$   or $Br(J/\psi\rightarrow\gamma f_0(1770)) = (3.7\pm 0.8)\cdot 10^{-3} \,$ is found.
This suggests a  significant glueball component in this scalar state.

\section{Acknowledgments}\label{sec9_acknow}

This work was done with the use of the IHEP (Protvino) Central Linux Cluster. 
The work is partially supported with the RFBR grant 20-02-00246. 

\bibliography{om_phi}

\end{document}